\documentclass[12pt, floatfix]{iopart}

\usepackage{graphicx}
\usepackage{dcolumn}
\usepackage{bm}
\usepackage{subfigure}
\usepackage{amssymb}
\usepackage{float}
\usepackage{multirow}

\newcolumntype{.}{D{.}{.}{4}}
\newcolumntype{p}{D{(}{(}{2}}

\renewcommand{\thesubfigure}{(\roman{subfigure})}
\makeatletter
\renewcommand{\@thesubfigure}{\thesubfigure\space}
\renewcommand{\p@subfigure}{\thefigure}
\makeatother
\def\bdm{\begin{displaymath}}
\def\edm{\end{displaymath}}

\begin{document}

\title{Constraining Inflation}
\author
{Peter Adshead and Richard Easther}
\address
{
Department of Physics, Yale University, New Haven,  CT 06520, USA}

\date{\today}

\begin{abstract}
Slow roll reconstruction is derived from the Hamilton-Jacobi
formulation of inflationary dynamics. It automatically includes
information from sub-leading terms in  slow roll, and facilitates
the inclusion of priors based on the duration on inflation.   We
show that at low  inflationary scales  the Hamilton-Jacobi equations
simplify considerably.  We provide a new classification scheme for
inflationary models, based solely on the number of parameters needed
to specify the potential, and provide forecasts for likely bounds on
the slow roll parameters   from future datasets.    A minimal
running of the spectral index, induced solely by the first two slow
roll parameters ($\epsilon$ and $\eta$) appears to be  effectively
undetectable by realistic Cosmic  Microwave Background  experiments.
However, we show that the ability to detect this signal  increases
with the  lever arm in comoving wavenumber, and we conjecture that
high redshift 21 cm data may allow tests of second order consistency
conditions on inflation.  Finally, we point out that the second
order corrections  to the spectral index are correlated with the
inflationary scale, and thus the amplitude of the CMB B-mode.

\end{abstract}

\maketitle

\section{Introduction}

Inflation is an elegant explanation for the large scale appearance
of our universe. Causally connected regions of space  are swept
outside the Hubble horizon during a phase of accelerated expansion
and cross back during a later epoch of regular expansion.  The
inflationary predictions of a flat universe, along with a Gaussian,
adiabatic and nearly scale invariant power spectrum are perfectly
consistent with current data. Moreover,  the observed
anti-correlation between temperature and polarization at large
scales provides further support for the inflationary origin of the
initial density perturbations.

While inflation  predicts the overall form of our visible universe,
we have very little understanding of the physical mechanism that
generates the accelerated expansion.  Consequently, there is
considerable interest in ``reverse engineering'' the inflationary
potential from astrophysical data.  The primordial power spectrum
can be written as a function of the slow roll
parameters.  In their simplest form, the slow roll parameters are
expressed as derivatives of the potential, so measuring
the spectrum to arbitrary precision would yield a
Taylor expansion of the potential.
Unfortunately,  the simplicity of this scheme is undermined by the
practical challenges that arise during its implementation.  Recall
that the evidence for {\em any\/} scale dependence in the primordial
spectrum is still preliminary, although there is indeed growing
evidence the spectrum is {\em red\/}, or has diminishing power at
short scales~\cite{Tegmark:2006az}. Reconstructing the potential
requires reliably distinguishing  between models which {\em
all} predict slightly different red spectra, which would need an improvement in the accuracy of  cosmological parameter determinations  by a further order of magnitude. Given both cosmic variance and the practical challenges of foreground subtraction it is not clear that this is possible, even in principle.  This is a particularly
pressing problem if we are limited to data derived from the CMB
[Cosmic Microwave Background] and LSS [Large Scale Structure]. These
two sources of data are sensitive to a range of scales that differ by a total factor of perhaps $10^4$ but in simple
models of inflation the primordial spectrum changes very slowly, undermining the power of any reconstruction program.     Despite these challenges, there are
powerful motivations for theoretical studies of reconstruction.  The
first is that the physical basis of inflation (if inflation is, in fact,
the source of the primordial fluctuations) is one of the most
important open questions in all of cosmology, and answering it is
likely to shed light upon particle physics at very high energies.
Secondly, a number of proposed experiments aim to measure the
fundamental spectrum to very high precision; we wish to determine
their ability  to constrain the overall inflationary parameter
space, even though many of them will take decades rather than years
to implement.

The reconstruction of the inflationary potential was was first
discussed in the early 1990s
\cite{Turner:1993su,Copeland:1993jj,Copeland:1993ie,Liddle:1994cr}.
This paper builds on the slow roll reconstruction algorithm,
proposed and implemented by Easther and Peiris
\cite{Peiris:2006sj,Peiris:2006ug,Easther:2006tv}, which grew out of
a Monte Carlo approach based on the inflationary flow equations
\cite{Hoffman:2000ue,Kinney:2002qn,Easther:2002rw,Kinney:2006qm,Powell:2007gu}.
Originally reconstruction was based on taking the measured values of
the spectral indices (and their running), solving for the slow roll
parameters, and deducing the form of the potential
\cite{Lidsey:1995np}. Since we are interested in the {\em
inflationary\/} parameter space, there is no need to compute the
spectral indices, as these  have no fundamental significance.
Rather, one can include the slow roll variables directly in the
cosmological parameter set and bound them using Monte Carlo Markov
Chain fits to cosmological data \cite{Liddle:2003as,Peiris:2006sj,
Peiris:2006ug}.   Slow roll reconstruction is thus an  optimal approach to
recovering the inflationary potential from data, since it makes use
of all available information. Moreover, while it makes use of the
slow roll \emph{expansion}, we need never employ the slow roll
\emph{approximation}, as the truncated Hamilton-Jacobi hierarchy can
be solved exactly \cite{Liddle:2003py}. Slow roll reconstruction
thus captures all the correlations between the slow roll parameters,
including those at second and higher order.\footnote{As explicitly
implemented in \cite{Peiris:2006sj,Peiris:2006ug}, slow roll
reconstruction uses approximate expressions for the perturbation
spectrum, but this is simply a matter of convenience
\cite{Peiris:2006sj}.   One can always solve the perturbation mode
equations numerically, and this approach has been
explored \cite{Lesgourgues:2007aa,Hamann:2008pb}. As we will see,
computing the spectrum with the {\em scale dependent\/} slow roll
parameters yields a very good match to the exact calculation.
Numerically evaluated spectra eliminate a source of uncertainty in
the parameter constraints, but are not justified by the quality of
presently available data.}

A number of other approaches to reconstruction have been proposed.
In particular,  Leach and collaborators write the spectral indices
(and their running) in terms of the slow roll parameters at a fixed
pivot \cite{Leach:2002ar,Leach:2002dw, Leach:2003us}. When used with data which probes a small range of scales, this
approach is functionally identical to slow roll reconstruction,
since the slow roll parameters are effectively constant.  However,
we will show that as the lever arm in wavelength becomes large, the
scale dependence of these parameters can be significant. In
principle, one could account for this running by computing the
perturbation spectrum using expressions that include higher order
corrections in slow roll, but this approach will become
algebraically cumbersome at some point.  Conversely,
\cite{Malquarti:2003ia} explores constraints on the slow roll
parameters imposed by demanding a sufficient period of inflation. Finally Cline and Hoi look at reconstructing inflationary models with significant running within the Hamilton-Jacobi formalism \cite{Cline:2006db}.
Since slow roll reconstruction implicitly predicts the form of the
potential it naturally makes use of this information, and one can
restrict fits to parameter values that allow a sufficiently long
period of inflation.

The aim of this paper is to forecast the parameter constraints we
can expect from slow roll reconstruction  when it is applied to
future datasets, and to explore the differences between slow roll
reconstruction and fits to the usual spectral
parameters. In the process, we show that low scale inflation (i.e.
energies significantly below $10^{15}$~GeV) is described by a
simplified set of Hamilton-Jacobi equations, as the first parameter
($\epsilon$) is effectively absent from the dynamical system. We
present a new classification scheme for inflationary models based on
the number of free parameters needed to specify the potential --
rather than its shape -- and show  this is naturally related to the
(truncated) slow roll hierarchy.  We then use Fisher matrix
calculations \cite{Fisher:1935,Tegmark:1996bz}  to explore how
constraints on the inflationary parameter space improve with the
lever arm in wavelength probed by the dataset.\footnote{We utilize
\texttt{CAMB}, {http://camb.info/}, \cite{Lewis:1999bs} for
calculations of the CMB spectra and the matter power spectrum, and
the two sided derivative methodology outlined in
\cite{Eisenstein:1998hr}.} With a purely Gaussian likelihood
function (which is an assumption of the Fisher matrix formalism) the
CMB alone does not probe a large enough range of wavelengths   to
produce results that differ significantly from those obtained via
fits to the standard spectral parameters. However, if one has access
to the primordial spectrum at very short scales (e.g. via high
redshift 21 cm measurements), terms that are second order in slow
roll may become significant.  In practice, the likelihood is far
from Gaussian, so this analysis is effectively a worst case scenario. In particular, slow roll reconstruction allows us to include
information about the {\em duration} of inflation into parameter
estimates; these constraints are not captured by a Fisher matrix
analysis. Models with a significant tensor spectrum generically
require a longer period of inflation than those with an unobservable
tensor component; we show how this information -- along with a very
mild prior on the post-inflationary equation  of state -- can give
further leverage to slow roll reconstruction.

Since we are considering constraints on the inflationary parameter
space, we are primarily interested in models for which
$\Omega_{\mbox{\tiny Tot}} \equiv 1$; the flatness of the spatial
hypersurfaces is a key prediction of inflation.  This is an implicit
{\em assumption\/} of slow roll reconstruction, since the
Hamilton-Jacobi equations are derived after ignoring the spatial
curvature term in the Einstein equations.   If $\Omega_{\mbox{\tiny
Tot}}$ differs slightly from unity today the longest modes will have
left the horizon just as inflation began.  Consequently, slow roll
reconstruction (as currently implemented) is only  self-consistent
when inflation lasts long enough to suppress any transients
associated with the pre-inflationary initial conditions {\em
before\/}  modes which contribute to the quadrupole leave the
horizon.   We also assume that the entire primordial perturbation
spectrum was generated during inflation, and contains no significant
contribution from cosmic strings, or other non-inflationary mechanisms. Finally, we treat
the dark energy as a pure cosmological constant,  although relaxing
this assumption would not significantly modify our key conclusions.

The CMB temperature peak morphology is  very well understood
\cite{Bond:1993fb, Knox:1994qj, Jungman:1995bz, Zaldarriaga:1998ar,
Hu:2001bc}, along with the importance of the $E$ and $B$
polarization modes. The $\ell$th multipole of the CMB anisotropy
corresponds to a wavenumber $k^{-1}\simeq 2/(H_{0}\ell)$, so the CMB
probes scales from $k\sim 2\times10^{-4}h~ $Mpc$^{-1}$ at the
quadrupole to $k\sim 0.18 h~ $Mpc$^{-1}$ at $\ell \sim 1500$. Beyond
$\ell\sim 1500$, the primordial $C_\ell$ decay sharply due to Silk
damping, which reflects the finite width of  the surface of last
scattering.  It is  thus difficult to measure the primordial  CMB at
shorter scales. Moreover, foregrounds and secondary anisotropies
typically grow in amplitude at smaller  scales.  In what follows,
we will consider perfect CMB measurements out to
$\ell_{\mbox{\tiny MAX}}$ of up to 2,500 in order to explore the
lever arm yielded by data over a large range of angular scales.
However, given the difficulties associated with foreground
subtraction, these calculations are essentially gedanken
experiments.

On the other hand, we are not limited to CMB information alone. The
information contained in LSS data is largely orthogonal to that in
the CMB, breaking many parameter degeneracies.  This `cosmic
complementarity'  is well known \cite{Zaldarriaga:1996xe,
Bond:1997wr, Eisenstein:1998hr, Wang:1998gb}, and the slow roll reconstruction
can obviously make use of this data.   While the linear regime of
structure formation is very well understood, recovering the
primordial spectrum at length scales which have undergone nonlinear
evolution is a challenging task. The smallest scale still in the
linear regime probed by low low-redshift LSS experiments is not
wildly different from that probed by high resolution measurements of
the CMB. In what follows, we are primarily interested in the {\em
differences\/} between slow roll reconstruction and analyses based
on the usual spectral variables, and adding LSS data would improve
{\em both\/} approaches. Furthermore, forecasts of parameter
uncertainties for LSS data depend significantly on the design of the
experiment.  Consequently, in this analysis we have focussed on CMB
data alone. One can probe smaller scales by looking at very high
redshift data, when these modes were still in the linear regime.
Options in this area include very deep galaxy surveys, or
Lyman-$\alpha$ experiments.  More speculatively -- but with far
greater potential power -- the high redshift 21 cm background may
allow probes of the primordial spectrum at very small wavelengths
\cite{McQuinn:2005hk}.  We plan to look at constraints derived from
combined fits to CMB and 21 cm data in a future paper, but the
calculations here suggest that such a measurement might permit tests
of higher order inflationary consistency conditions
\cite{Copeland:1993zn,Kosowsky:1995aa,Lidsey:1995np}. Finally, a BBO style experiment is sensitive
to the primordial {\em tensor} spectrum at solar system scales, and
slow roll reconstruction could   take advantage of this data to put
exquisitely accurate constraints on the inflationary potential.

This paper is organized as follows. In \S\ref{inflandpower} we
review inflationary dynamics, the slow roll approximation and our
construction of the primordial spectrum. In  \S\ref{loweps} we
consider analytic models with low inflationary scales,  and
construct a new classification scheme for inflationary models in
\S\ref{classificationsection}. In \S\ref{fisherintro} we review the
Fisher matrix formalism for CMB anisotropy experiments, and describe
the resulting forecasts in \S\ref{results}. We conclude in
\S\ref{concl}. Finally, in an Appendix we give a more detailed
analysis of the slow roll dynamics at low inflationary scales.

\section{Inflation and the Primordial Power Spectrum}\label{inflandpower}

\subsection{The Background}

We use the Hamilton-Jacobi formulation of inflationary dynamics,
expressing the Hubble parameter as a function of
$\phi$, rather than as a function of time. Thus $H\equiv
H(\phi)$, and we assume that $\phi$ is  monotonic.
The equations of motion are
\cite{Lidsey:1995np, Grishchuk:1988,Muslimov:1990be,
Salopek:1990jq,Salopek:1990re,Lidsey:1991zp}
\begin{equation}\label{dotphi}
\dot{\phi}  =  -\frac{m_{\rm{Pl}}^{2}}{4\pi}H'(\phi),
\end{equation}
\begin{equation}\label{Hamilton-Jacobi}
[H'(\phi)]^{2}-\frac{12\pi}{m_{\rm{Pl}}^{2}}H^{2}(\phi) =
-\frac{32\pi^{2}}{m_{\rm{Pl}}^{4}}V(\phi).
\end{equation}
Primes denote derivatives with respect to the field, while an
overdot denotes derivatives with respect to coordinate time.
Equation (\ref{Hamilton-Jacobi}) is the Hamilton-Jacobi equation,
and describes inflation in terms of the Hubble parameter, $H(\phi)$,
rather than the potential, $V(\phi)$. The Hubble parameter, being a
geometric quantity, describes the spacetime dynamics, whereas
particle physics constructions predict  $V(\phi)$.   We can thus
discuss slow roll inflation without specifying the  particle physics
that generates inflation.   The HSR [Hubble slow roll] parameters
$^{\ell}\lambda_{H}$ are defined by the infinite hierarchy of
differential equations \cite{Kinney:2002qn}:
\begin{eqnarray}\label{epsphi}
\epsilon(\phi) & \equiv &
\frac{m_{\rm{Pl}}^{2}}{4\pi}\left[\frac{H'(\phi)}{H(\phi)}\right]^{2},\\
^{\ell}\lambda_{H} & \equiv &
\left(\frac{m_{\rm{Pl}}^{2}}{4\pi}\right)^{\ell}\frac{(H')^{\ell-1}}{H^{\ell}}
\frac{d^{\ell+1}H}{d\phi^{(\ell+1)}};\;\ell \geq 1.
\end{eqnarray}
The usual slow roll parameters are $\eta = {}^{1}\lambda_{H}$ and
$\xi = {}^{2}\lambda_{H}$. If we truncate the hierarchy,  so that
${}^{\ell}\lambda_{H} =0 $ for all $\ell > M$ at some $\phi_0$,
then these ${}^{\ell}\lambda_{H}$ vanish everywhere. When truncated at
order $M$, the hierarchy can be solved exactly to obtain
\cite{Liddle:2003py}
\begin{equation}
\frac{H(\phi)}{H_{0}} = \sum_{n =
0}^{M+1}B_{n}\left(\frac{\phi}{m_{\rm{Pl}}}\right)^{n},
\end{equation}
where the $B_{n}$ are specified by the initial values of the HSR
parameters:
\begin{equation}
B_{0} = 1,\quad B_{1} = \sqrt{4\pi\epsilon_{0}},\quad B_{\ell+1} =
\frac{(4\pi)^{\ell}}{(\ell+1)!B_{1}^{\ell-1}}{}^{\ell}\lambda_{H,\,0}.
\end{equation}
The subscript $0$ refers to their value at the moment the fiducial
mode $k_{0}$ leaves the horizon (when $aH = k_{0}$) and $\phi =
\phi_{0} = 0$. In this analysis we set $k_{0} = 0.05$ Mpc$^{-1}$, which
corresponds to $\ell\sim500$ in the CMB.

Substituting equation (\ref{epsphi}) into equation
(\ref{Hamilton-Jacobi}) gives the potential
\begin{equation}
V(\phi) =
\frac{3m_{\rm{Pl}}^{4}}{8\pi^{2}}H^{2}(\phi)\left[1-\frac{1}{3}
\epsilon(\phi)\right],
\end{equation}
while $N$, the number of e-folds before the end of inflation is
\begin{equation}\label{dNdphi}
\frac{dN}{d\phi} = \frac{4\pi}{m_{\rm{Pl}}^{2}}\frac{H}{H'} =
\frac{2\sqrt{\pi}}{m_{\rm{Pl}}}\frac{1}{\sqrt{\epsilon(\phi)}},
\end{equation}
and $\phi$ and $k$ are related by
\begin{equation}\label{dphidlnk}
\frac{d\phi}{d\ln k} =
-\frac{m_{\rm{Pl}}}{2\sqrt{\pi}}\frac{\sqrt{\epsilon}}{1-\epsilon}.
\end{equation}
The scale dependence of the slow roll parameters follows from equations (\ref{epsphi}) and (\ref{dNdphi}):
\begin{eqnarray} \label{flow1}
\frac{d\epsilon}{dN} & = & 2\epsilon(\eta-\epsilon),\\
\frac{d\eta}{dN} & = & -\epsilon\eta+ \xi, \\ \label{flowellfull}
\frac{d^{\ell}\lambda_{H}}{dN} & = &
[(\ell-1)\eta-\ell\epsilon]^{\ell}\lambda_{H}+{}^{\ell+1}\lambda_{H},
\end{eqnarray}
and the truncation property mentioned above can be derived from the
last equation.  Note that these formulae use $N$ as their
independent variable.

Once we have specified the values of the slow roll parameters at
some fiducial scale, we may ``flow'' to any other scale by using the
slow roll hierarchy.  The scale dependence of the
slow roll parameters ensures that any non-trivial correlation
between the slow roll parameters will only strictly apply at a
single value of $\phi$. For instance, if might assume the existence of a special point
where $\epsilon$ and $\eta$ were non-zero, while ${}^i\lambda_H=0$
for $i=3,\cdots,N-1$ and ${}^N\lambda_H\ne0$ (see
\cite{Ballesteros:2005eg} for an example).  In principle, one could   analyze this type of model  by adding $k_0$ to the
parameter set in order to marginalize over the model-dependence in
the mapping between $k$ and $\phi$.

\subsection{The Perturbations}

The  scalar and tensor perturbations obey \cite{Mukhanov:1990me},
\begin{equation}\label{uk}
\frac{d^{2}u_{k}}{d\tau^{2}} +
\left(k^{2}-\frac{1}{z}\frac{d^{2}z}{d\tau^{2}}\right)u_{k} = 0 \, ,
\end{equation}
\begin{equation}\label{vk}
\frac{d^{2}v_{k}}{d\tau^{2}} +
\left(k^{2}-\frac{1}{z}\frac{d^{2}z}{d\tau^{2}}\right)v_{k} = 0 \, .
\end{equation}
Here $\tau$ is conformal time, $u_{k}$ are the Fourier modes of the
gauge invariant Mukhanov potential describing the intrinsic
curvature perturbation, $v_{k}$ are the analogous Fourier modes
for the tensor perturbations, while $z = a\dot{\phi}/H$ for scalar
perturbations, and $z = a$ for tensor perturbations. The power spectra are
\begin{eqnarray}
\mathcal{P}_{\mathcal{R}} & = &
\frac{k^{3}}{2\pi^{2}}\left|\frac{u_{k}}{z} \right|^{2} \, ,\\
\mathcal{P}_{h} & = & \frac{32k^{3}}{\pi
m_{\rm{pl}}^{2}}\left|\frac{v_{k}}{a} \right|^{2}  \, .
\end{eqnarray}
A first order expansion about the exact solution for power law
inflation gives \cite{Stewart:1993bc, Peiris:2006sj, Peiris:2006ug}
\begin{eqnarray}\label{Pr}
\mathcal{P}_{\mathcal{R}} & = &
\frac{[1-(2C+1)\epsilon+C\eta]^{2}}{\pi\epsilon}\left.\left(\frac{H}{m_{\rm{Pl}}}\right)^{2}\right|_{k=aH},\\
\label{Ph}\mathcal{P}_{h} & = &
[1-(C+1)\epsilon]^{2}\frac{16}{\pi}\left.\left(\frac{H}{m_{\rm{Pl}}}\right)^{2}\right|_{k=aH},
\end{eqnarray}
where $C = -2+\ln 2+\gamma\approx-0.729$ and $\gamma$ is the
Euler-Mascheroni constant. The power spectrum is normalized at $k_{0}$ by
setting
\begin{equation}
A_{s} =
\frac{[1-(2C+1)\epsilon_{0}+C\eta_{0}]^{2}}{\pi\epsilon_{0}}\left(\frac{H_{0}}{m_{\rm{Pl}}}\right)^{2},
\end{equation}
where $\epsilon_{0}$, $\eta_{0}$ and $H_{0}$ are the values of these
parameters at $k_{0}$.

\begin{figure*}
\begin{center}
\resizebox{8cm}{!}{\includegraphics{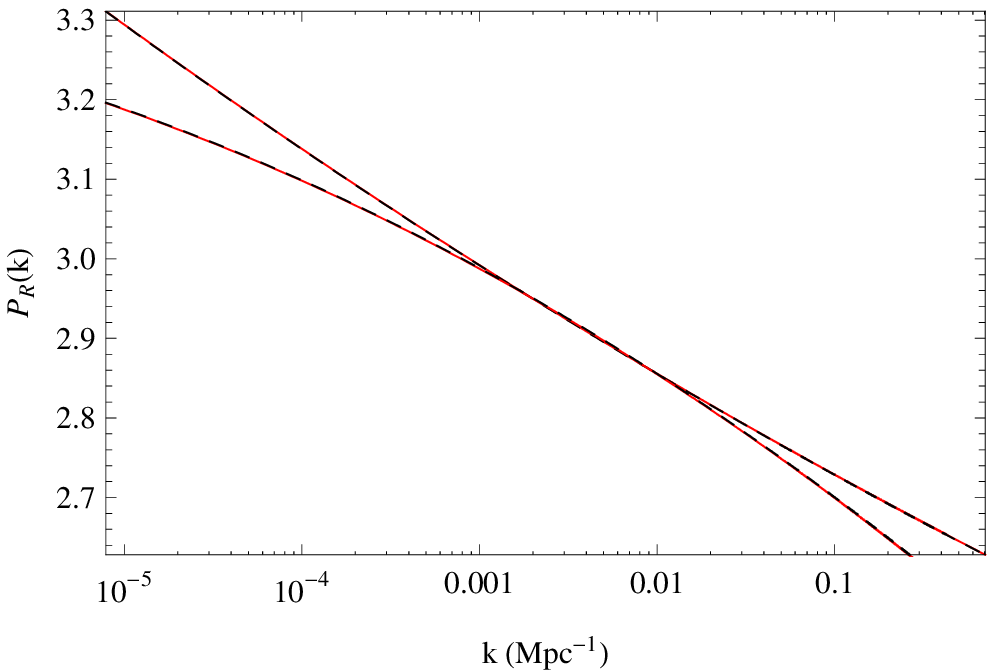}} \hspace{0.13cm}
\resizebox{8cm}{!}{\includegraphics{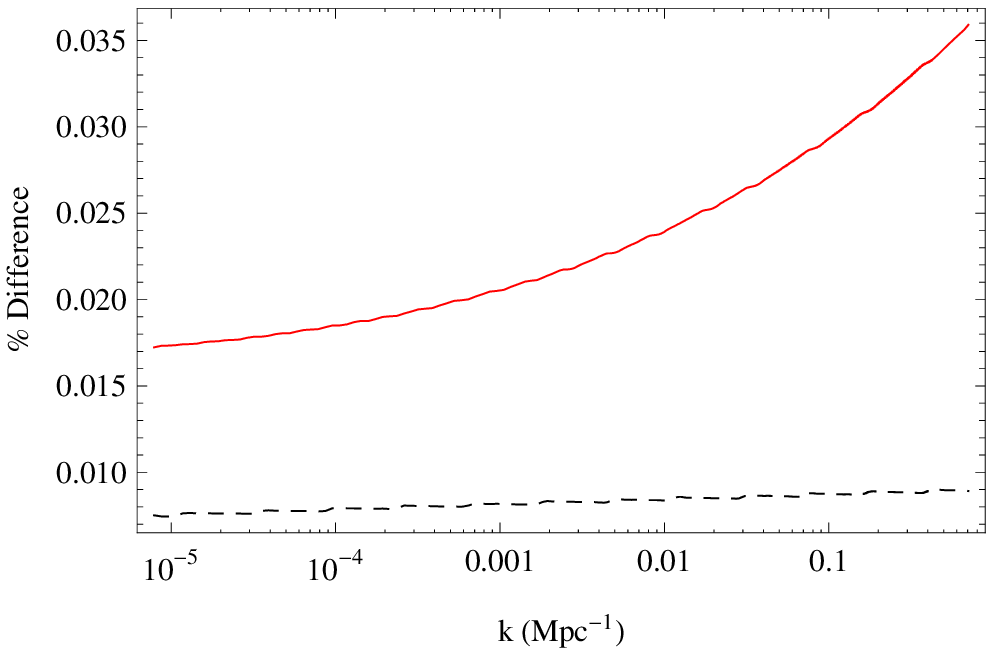}}
\end{center}
\caption{The top panel shows the primordial spectra for two
parameter choices; $\epsilon = 0.01$, $\eta = 0.01$, $\xi = 0$
(upper) and $\epsilon = 0.01$, $\eta = 0.01$, $\xi = 0.001$ (lower,
curved).   The bottom panel shows the difference between the
spectra, the dashed curve applies to $\xi = 0.0$ while the solid
curve corresponds to  $\xi = 0.001$.  For extreme values of $\xi$
these discrepancies can become large, but these models typically
have a very low number of e-folds.}\label{spectra}
\end{figure*}

One can always solve the perturbation evolution equations
numerically  \cite{Adams:2001vc,Ringeval:2007am,Salopek:1988qh,
Peiris:2003ff}. This approach is implemented in
\cite{Lesgourgues:2007aa,Hamann:2008pb} who parametrize $H(\phi)$ as
a polynomial of finite order -- which is identical to slow roll
reconstruction,\footnote{Note that
\cite{Lesgourgues:2007aa,Hamann:2008pb} also consider fits to a
polynomial expression for $H^2(\phi)$.  Hamann, Lesgourgues, and
Valkenburg  \cite{Hamann:2008pb} present constraints on the first
three slow roll parameters from WMAP3 \cite{Spergel:2006hy} and
ACBAR \cite{Reichardt:2008ay} derived with two different
approximations to the spectrum, as well as the numerically evaluated
mode equations. Their analysis shows that the difference between
these results is entirely explained by the implicit priors on the
duration of inflation, and not the accuracy with which $P(k)$ is
evaluated. The ability to include constraints based on the duration
of inflation is a key feature of slow roll reconstruction, and we
return to this topic below. } since the truncated flow hierarchy is
solved exactly by a polynomial in $H(\phi)$ \cite{Liddle:2003as}.
Since equation~(\ref{Pr}) uses the scale dependent slow roll
parameters -- which are matched to a value of $k$ by solving
equation~(\ref{dphidlnk}) -- it accurately tracks the exact spectrum
for the combinations of slow roll parameters one is likely to
encounter in practice.  We plot the difference between
equation~(\ref{Pr}) and the exact spectrum for two representative
sets of slow roll parameters in Figure~\ref{spectra}. In this paper,
we perform our calculations using equations~(\ref{Pr})  and
(\ref{Ph}), but a numerically computed spectrum would banish the
last vestiges of the slow roll {\em approximation} from this
analysis, and might be justified when dealing with very high quality
data, especially if the running in the spectral index turns out to
be non-trivial.

From (\ref{Pr}) and (\ref{Ph}), one can also recover expressions
for the usual spectral indices and their scale dependence
\cite{Lidsey:1995np}:
\begin{eqnarray}\label{ns}
n_{s} & = & 1+2\eta -
4\epsilon-2(1+\mathcal{C})\epsilon^2-\frac{1}{2}(3-5\mathcal{C})\epsilon\eta+\frac{1}{2}(3-\mathcal{C})\xi,
\\\label{r}
r & = & 16\epsilon[1+2C(\epsilon-\eta)],\\
\frac{dn_{s}}{d\ln k} & = & \nonumber
-\frac{1}{1-\epsilon}\left\{2\frac{d\eta}{d N}-4\frac{d\epsilon}{dN}
- 4(1+\mathcal{C})\epsilon\frac{d\epsilon}{dN} - \right.\\
& {} & \left. \label{running} \frac{1}{2}(3-5\mathcal{C})
\left(\epsilon\frac{d\eta}{dN}+\eta\frac{d\epsilon}{dN}\right)+\frac{1}{2}(3-\mathcal{C})\frac{d\xi}{dN}\right\},\\
n_{t} & = & \label{nt}
-2\epsilon-(3+\mathcal{C})\epsilon^{2}+(1+\mathcal{C})\epsilon\eta,
\end{eqnarray}
where $\mathcal{C} = 4(\gamma+\ln 2)-5$ and is not to be confused
with $C$ above. The expressions are not used during slow roll reconstruction,  but are employed when we make
comparisons to the empirical characterization.

\subsection{The Duration of Inflation}

We need make no {\em  explicit\/} assumption about the duration of
inflation.  Other HSR based analyses often pick a specific set of
parameters, then integrate forward in time to find the moment at
which inflation ends, and then work backwards from this point to
find the slow roll parameters some fixed number of e-folds before
the end of inflation
\cite{Kinney:2002qn,Easther:2002rw,Kinney:2006qm,Powell:2007gu}. By
running Markov Chains with the slow roll variables in the parameter
set we learn their values at the moment when cosmological scales
were leaving the horizon. We can then evolve them forwards in time,
to discover the point (if any) at which inflation ends, which occurs
when $\epsilon = 1$.   In three
parameter fits to current data, one finds considerable support for a
running index, which requires large values of $\xi$, leading to
models with $N\sim15$ or even less
\cite{Peiris:2006ug,Easther:2006tv}.  If we assume that we have
enough slow roll parameters to describe the inflationary potential
this situation is clearly not self-consistent.  One can posit  the
existence of a secondary period of inflation, or expand the
parameter set in order  to make $\xi$ strongly scale-dependent, but
either of these ``solutions'' implies that the set
$\{\epsilon,\eta,\xi\}$ is incomplete.   Conversely, if we find a
large or even unbounded number of e-folds for a given set of slow
roll parameters, the implication is that inflation ends via some
sort of hybrid or waterfall transition
\cite{Linde:1993cn,Copeland:1994vg}, where the field evolves in a
direction orthogonal to the original inflaton trajectory. These two
scenarios are analogous to following a path that leads down a
mountainside to a valley or walking along a gentle plain and then
encountering a sharp cliff, respectively. In the former case, the
end of inflation can be extrapolated from the slope of the potential
some distance away from the minimum, while in the latter case the
inflaton field receives little or no advance warning that inflation
is about to end.

The primordial tensor amplitude is set by the energy scale of
inflation. Assuming slow roll, the scalar amplitude is given by   \cite{Lyth:2000}
\begin{equation} \label{Pquick}
\mathcal{P}_{S,0} = \frac{8V_k}{3m_{\rm{Pl}}}\frac{1}{\epsilon},
\end{equation}
where $\mathcal{P}_{S,0}\simeq2.6\times10^{-9}$ \cite{Leach:2002dw}
and $V_k$ is the value of the potential as the mode $k$ leaves the
horizon. This energy scale can be translated into a constraint on
the number of e-folds needed to reproduce the observed universe
\cite{Liddle:2003as}
\begin{eqnarray}\label{Nefolds}
N(k) & = & 63.3+\frac{1}{4}\ln\epsilon(k)-
\ln\left[\frac{k}{a_{0}H_{0}}\right]+
\ln\left[\frac{V_{k}^{\frac{1}{4}}}{V_{end}^{\frac{1}{4}}}\right]
-\frac{1}{3}\ln\left[\frac{V_{end}^{\frac{1}{4}}}{\rho_{reh}^{\frac{1}{4}}}\right],
\end{eqnarray}
where $V_{end}$ is the energy scale at the end of inflation,
$\rho_{reh}$ is the energy density at reheating and $a_{0}H_{0}$ is
the expansion rate today. As pointed out by Kinney and Riotto
\cite{Kinney:2005in}, the undetermined parameters in $N(k)$ induce a
\textit{theoretical} uncertainty in associating inflationary
parameters with a potential.

By writing down equation~(\ref{Nefolds}) we have assumed that the
universe is matter dominated after the end of inflation and then
radiation dominated between the end of reheating and
matter-radiation equality, which amount to assuming that  the
effective equation of state parameter varies between $w=0$ and
$w=1/3$ between the of inflation and nucleosynthesis.  If we include
matter with $w>1/3$, such as a kination field \cite{Chung:2007vz},
$N(k)$ will increase.  The minimum possible value of $w$ in a
decelerating universe is $-1/3$, and if $-1/3 < w <0$, $N(k)$ can be
substantially reduced.  Note that the term involving
$V_{k}$ implicitly contains $\epsilon$. However, in slow roll
inflation $V_{k}$ is not expected to differ considerably from
$V_{end}$, and thus this term is not considered to be important.

Slow roll reconstruction allows one to include priors based on the
duration of inflation when fitting to the slow roll parameters.  In
particular, the duration of inflation depends very strongly on the
value of $\xi$  \cite{Easther:2006tv}, and even the modest
requirement that $N>30$ induces stringent bounds on the parameter
space   \cite{Peiris:2006sj}.  However, studying equations
(\ref{Pquick}) and (\ref{Nefolds}) shows that the value of $N$
depends strongly on $\epsilon$ -- while we can certainly have
inflationary models with $N\sim30$, this requires a tiny
inflationary scale, and thus a miniscule value of $\epsilon$.
Conversely, if $\epsilon \sim 0.01$, the inflationary scale must be
comparatively large, and $N \sim 50$. Consequently, we can implement
slow roll reconstruction with an $\epsilon$ dependent constraint on
the number of e-folds by using equation~(\ref{Pquick}) to write
$V_k$ as a function of $\epsilon$ and the  amplitude of the
primordial perturbation spectrum. This requires a mild prior on the
post-inflationary equation of state -- for instance that $w>0$
after the end of inflation.  This implicitly rules out exotic
scenarios such as a phase in which the universe is dominated by a
frustrated network of cosmic  strings \cite{Burgess:2005sb}, but
does not strongly constrain the post-inflationary universe.
Finally, a principal lesson of \cite{Hamann:2008pb} is that {\em
all\/} implementations of slow roll reconstruction make implicit
assumptions about the minimal permissible duration of inflation.
Consequently, slow roll reconstruction is most transparent when an
explicit prior on $N$ is included in the chains, even if that bound
is very mild, and we will pursue this topic in a separate paper.

\section{Slow Roll in the Low-$\epsilon$ limit}\label{loweps}

A primordial gravitational wave spectrum is often described as the
``smoking gun'' of inflation, since there is no credible alternative
mechanism for generating long wavelength gravitational waves.
Detecting this signal would determine the overall energy scale of
inflation since ${\cal P}_h \sim H^2$. Intriguingly, string
theoretic inflationary scenarios generically predict a negligible
amount of primordial gravitational radiation
\cite{McAllister:2007bg}, or $r< 10^{-10}$. In this  case,
$\epsilon$ must also be tiny, since $r\sim16\epsilon$. Conversely,
models with algebraically simple potentials typically have $r>
0.001$ \cite{Boyle:2005ug}.  The constraint on stringy models
can be understood in terms of the Lyth bound \cite{Lyth:1996im},
which requires that the inflaton's total excursion be  sub-Planckian
to ensure that its potential is not dominated by contributions from
higher-order operators.

The current observational bound on $r$ is relatively weak, $r<0.3$
\cite{Spergel:2006hy}, but upcoming CMB experiments may push this
value down to $r \sim 0.01$
\cite{Verde:2005ff,Taylor:2006jw,Bock:2006yf}. However, if $r$ is
many orders of magnitude smaller than unity it may be unobservable
by any conceivable experiment \cite{Knox:2002pe}. While $\epsilon$
must be tiny in these models, there is no equivalent constraint  on
$\eta$, and in many supergravity or stringy models the challenge is
to ensure that $|\eta| < 1$.  Consequently, we assume $1\gg
|\eta| \gg \epsilon$, and write the flow equations as
\begin{eqnarray} \label{flow2}
\frac{d\eta}{dN} &=& \xi, \\ \label{flowxi}
\frac{d\xi}{dN} &=& \eta\xi + {}^{3}\lambda_{H}, \\
\label{flowell}
\frac{d{}^{\ell}\lambda_{H}}{dN} &= &
(\ell-1)\eta{}^{\ell}\lambda_{H}+{}^{\ell+1}\lambda_{H}.
\end{eqnarray}
Here $\xi$ drives the evolution of $\eta$ and $^{3}\lambda_{H}$ drives
the evolution of $\xi$ (if we could measure it), and one can still
truncate the hierarchy.  We can also estimate $\epsilon$,
\begin{equation}
\frac{d\epsilon}{dN} = 2 \epsilon \eta + {\mathcal{O}}(\epsilon^{2})
,
\end{equation}
which can be solved to give
\begin{equation}
\epsilon(N) = \epsilon(N_0) \exp{\left[2 \int_{N_0}^N \eta(N) dN \right]}  .
\end{equation}
Since $\eta\ll1$, if $\epsilon$ is initially small, it stays small
and then rises super-exponentially as $\eta \sim 1$. Conversely, if
$\eta$ is also vanishingly small, we cannot usefully extend this
truncation, since $\epsilon$ is unique in that both of its source
terms in the slow roll hierarchy are suppressed when it is small. To
illustrate, suppose that the first $m-1$ slow roll parameters are
close to zero, but $^{m}\lambda_{H}\neq0$. However,
$^{m}\lambda_{H}$ generates $^{n}\lambda_{H}\neq0$ for all $n<m$, so
this condition amounts to choosing the initial conditions at a very
special point in the potential. The  coupling between
$^{m}\lambda_{H}$ and $^{m-1}\lambda_{H}$ means that after one
e-fold $^{m-1}\lambda_{H}$ will be at least as large as
$^{m}\lambda_{H}$. One may proceed inductively to see that after $m$
e-foldings (and probably sooner) all of the slow roll parameters
except for $\epsilon$ will differ significantly from zero.

When we drop $\epsilon$, we can solve the slow roll hierarchy
exactly if it is truncated at a relatively low order. These
solutions are explored in  \ref{lowscale}. Consider the specific
case  $\xi = 0$, $\eta\ll1$  and $\epsilon \ll \eta$, so $d\eta/dN
\approx 0$ and $\eta(N) \approx \eta(0)\equiv \eta_{0}$, which
corresponds to the ``Low-$\epsilon$ 1-Parameter'' model defined in
the next section. From equation (\ref{running}) it follows that
$n_s$ is effectively constant over cosmological scales. Equation
(\ref{flow1}) has the solution
\begin{equation}
\epsilon(N) =
\frac{\eta_{0}\epsilon_0}{\epsilon_0+(\eta_{0}-\epsilon_0)e^{-2\eta_{0}N}},
\end{equation}
where we have set $N=0$ at the fiducial scale for convenience. In
this approximation, we can solve exactly for where inflation ends:
\begin{equation}
N(\epsilon = 1) =
-\frac{1}{2\eta_{0}}\ln\left(\frac{\epsilon_0(\eta_{0}-1)}{\eta_{0}-\epsilon_0}\right),
\end{equation}
and we recall that unless $\eta_0<0$, inflation will continue indefinitely in this scenario.
\begin{figure*}[!t]
\begin{center}
\resizebox{11cm}{!}{\includegraphics{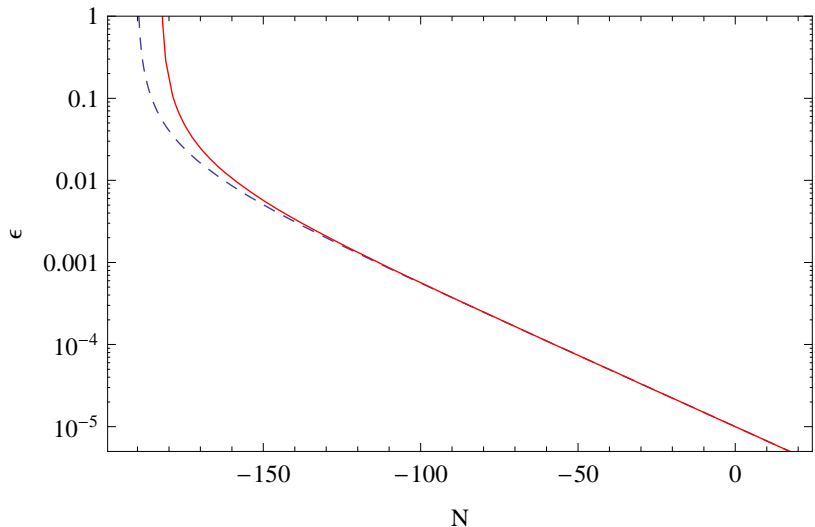}} \hspace{0.13cm}
\end{center}
\caption{Slow roll hierarchy truncated at $\xi=0$. The  plot shows
the evolution of $\epsilon$, the red curve is the exact result, the
dashed blue curve is the approximation with $\eta$ constant.   We
have set $\eta=-0.02$ and $\epsilon=10^{-5}$.}
\label{fig:efoldloweps}
\end{figure*}

If we have $\epsilon\sim10^{-10}$, and $\eta = -0.02$ (e.g. $n_s \approx 0.96$), we have inflation
ending $N\sim500$ e-folds after the fiducial scale has left the
horizon while $\epsilon\sim10^{-5}$ gives a futher $N\sim190$. As
illustrated by Figure \ref{fig:efoldloweps}, the approximate solution is
typically good up until the last couple of e-folds and leads to a slight overestimate of $N$.
Consequently, if $\eta \approx -0.02$, and $\epsilon$ is very small,
inflation either ends via a hybrid transition when it encounters an
abrupt cliff in the potential, or at least one higher order slow
roll parameter has a non-trivial value.
 Further, looking at equation (\ref{running}), if $\epsilon$ is tiny
and $^{\ell}\lambda_{H} = 0$ for $\ell> 2$ the running is entirely
dominated by $\xi$ and is thus negligible if $\xi = 0$.
Conversely, if $\epsilon \approx 0.01$, some running will be generated
by the $\epsilon^2$ term in the slow roll hierarchy, even if $\xi =
0$.

We also see that $\xi$ quickly dominates the dynamics when it is of
the same order of magnitude as $\eta$, as happens when we try to
match the central value of the running derived from  WMAP data. As
noted by \cite{Easther:2006tv} the effect of this is to reduce the
number of e-folds to an unacceptably small level.  If the running
is significant, then we would need at least one further slow-roll
parameter in order to ensure that $\xi$ itself is scale dependent.
Unless such a scenario is a prediction of a well-motivated model,
this situation would justifiably be regarded as a fine-tuning.

\section{Parameter Counting and Model Classification}\label{classificationsection}

We now describe a classification scheme for inflationary models,
based on the slow roll hierarchy.   In the past, it has been tacitly
assumed that  $\epsilon$ is the one parameter that {\em cannot\/} be
dropped from the slow roll hierarchy. Consequently, all previous
implementations of slow roll reconstruction have allowed for a
tensor component in the CMB, whereas the minimal $\Lambda$CDM
parameter set includes only scalar perturbations.  In the language
of slow roll, the $\Lambda$CDM parameter set is thus equivalent to
assuming that $\epsilon \lesssim 10^{-5}$, in which case $\epsilon$ plays no direct role in
determining the observable properties of the present universe, and we need to include it in our chains. We
thus propose the following scheme:
\begin{itemize}
\item Low-$\epsilon$, $N$-parameter.  The tensor-scalar ratio is assumed to be immeasurably small, and we truncate the slow roll hierarchy at ${}^{N+1}\lambda$.

\item High-$\epsilon$, $N$-parameter. The tensor-scalar ratio can be measurably different from zero,   and we truncate the slow roll hierarchy at ${}^{N}\lambda$.
\end{itemize}
This is distinct from the ``large field / small field / hybrid"
zoology (see e.g. \cite{Kinney:2006qm}).  Models with a large
variation in $\phi$ necessarily have a non-trivial $\epsilon$
\cite{Lyth:1996im,Easther:2006qu}, whereas the small field and
hybrid cases are distinguished by the sign of $\eta$. However, when
$\xi$ is nontrivial, $\eta$ can change sign during the course of
inflation, so ``shape-based'' taxonomies are best avoided with
general potentials.  Note that this classification does not specify
how inflation ends, whether via a hybrid transition, or a violation
of slow roll -- the valley or the cliff.

\begin{table}
\centering\label{classification}
\begin{tabular}{|c|c|c|c|c|}
\hline
Class & HSR & Spectral & Shorthand \\ \hline
High-$\epsilon$ 1-Parameter  & $\epsilon$ & $n_{s}-1 = \frac{1}{4}r = -2n_{t}$, $\alpha\approx0$  & -- \\ \hline
Low-$\epsilon$ 1-Parameter   &   $\eta$ & $n_{s}\neq1$, $r\approx0$, $\alpha\approx0$ & $\Lambda$CDM   \\ \hline
High-$\epsilon$ 2-Parameter   & $\epsilon$, $\eta$ & $n_{s}-1 \neq \frac{1}{4} r= -\frac{1}{2}n_{t}$, $\alpha\approx0$ & $\Lambda$CDM+$r$ \\ \hline
Low $\epsilon$  2-Parameter  &  $\eta$, $\xi$ & $n_{s}\neq1$, $r\approx0$, $\alpha\neq0$ & $\Lambda$CDM+$\alpha$ \\   \hline
High-$\epsilon$ 3-Parameter  & $\epsilon$, $\eta$, $\xi$ & $n_{s}-1 = \frac{1}{4}r = -\frac{1}{2}n_{t}$, $\alpha\neq0$ & $\Lambda$CDM+$r$+$\alpha$   \\ \hline
Low-$\epsilon$ 3-Parameter   &   $\eta$, $\xi$, ${}^{3}\lambda_{H}$ &
$n_{s}\neq1$, $r\approx0$, $\alpha\neq0$ ... & --\\
\hline
\end{tabular}
\caption{We show the non-trivial variables for models with up to
three HSR parameters, and the equivalent set of variables written in
terms of the scalar and tensor indices. The final column denotes the
corresponding shorthand for the resulting model. The canonical
$\Lambda$CDM case corresponds to assuming that $\epsilon \ll
|\eta|$. In this column $r$ refers to adding a tensor spectrum,
while $\alpha$ denotes the running of the spectral index.  }
\label{classes}
\end{table}

Within this schema, we can consider $N$ parameter models. A
Low-$\epsilon$ 1-Parameter model is almost entirely equivalent to
standard $\Lambda$CDM, since $\eta$ is the only nontrivial slow roll
parameter.  In this case the running in $\eta$ is effectively zero,
which leads to a constant spectral index, with no tensor component.
Conversely, a  High-$\epsilon$ 1-Parameter model has no obvious
analog, since  $n_{s}$, $r$, $n_{t}$ and $\alpha$ are \emph{all}
specified in terms of a single parameter. Physically, setting $\eta
=0$ means that $H(\phi)$ is a pure quadratic function.  Since
$\epsilon$  increases with time, a very small value of $\epsilon$ at
CMB scales would imply that inflation must have ended via a sudden
hybrid-style transition.

With two parameters, a High-$\epsilon$ model corresponds to
$\Lambda$CDM+$r$, where the tensor spectrum obviously has an
inflationary prior that links the amplitude and spectral index.
Moreover, as we will see later, the flow equations imply a weak
scale dependence in $\epsilon$ and $\eta$. Thus, the running of the
effective spectral index is non-zero -- and for very high quality
data, this increases the leverage we can obtain from slow roll
reconstruction.  Conversely, a Low-$\epsilon$, two parameter model
corresponds to  $\Lambda$CDM+$\alpha$, but we now must be careful to
ensure that for comparatively large and positive values of $\xi$ the
total duration inflation is self-consistent.  If we do wish to
consider a large running at CMB scales, our one recourse is to add
$^{3}\lambda_{H}$  to our parameter set, leading to either a
High-$\epsilon$ 4-Parameter  model, or a Low-$\epsilon$ 3-Parameter
model.  In this case the tilt is transient, but we are left with a
potential that has several free parameters. The classification
scheme is summarized in  Table \ref{classification}.

As we noted above, this analysis does not specify the mechanism that
ends inflation -- that is, whether we are rolling toward a cliff or
a valley.  However, since slow roll reconstruction effectively
specifies the overall form of the potential, we can add further cuts
to our parameter space by insisting on a sufficient overall duration
of inflation, as described at the end of Section~2. This is not
possible when using the spectral indices and amplitudes, since these
variables contain no information about the duration of inflation.

\section{Error Estimation and Fisher Information}\label{fisherintro}

For our error forecasts we make use of the Fisher information matrix
\cite{Fisher:1935}, a measure of the width and shape of the
likelihood function around its maximum,
\begin{equation}
F_{ij} = -\left\langle \left.\frac{\partial^{2} L}{\partial
\alpha_{i}\alpha_{j}}\right\rangle\right|_{\alpha = \bar{\alpha}},
\end{equation}
where $L\equiv \ln\mathcal{L}$ and the $\alpha_{i}$ denote model
parameters. The Cramer-Rao inequality then says that the minimum
possible standard deviation on a single parameter, $\alpha_{i}$,
estimated from the data is $1/\sqrt{F_{ii}}$. This minimum standard
deviation rises to $1/\sqrt{(F^{-1})_{ii}}$, if all parameters are
estimated from the same data.  Previous treatments of this topic
include \cite{Kinney:1998md,Verde:2005ff}, and we follow the
formalism laid out in  \cite{Verde:2005ff}.

We restrict our attention to  CMB data -- in this paper, our
principal concern  is to compare slow roll reconstruction to
fits to the spectral variables. Adding more information further
constrains the free parameters, and thus accentuates the advantages
the HSR formulation. Moreover, we are not considering the impact of
priors based on the duration of inflation (which puts sharp cuts on
the allowed region of parameter space), so what follows is
essentially a worst case analysis for slow roll reconstruction.

Observations of the CMB measure the polarization and the anisotropy
of the temperature of the radiation
in terms of spherical harmonics, from which we obtain the $C_\ell$ for each of the spectra
$C_{T\ell}$, $C_{E\ell}$, and $C_{B\ell}$ and the cross-correlation
$C_{C\ell}$.  Assuming that the CMB multipoles are Gaussian
distributed and letting $\bf{\alpha}$ denote our vector of parameters,
with $\bar{\bf{\alpha}}$ the fiducial values, the Fisher matrix for
a temperature/polarization measurement can be written
\begin{equation}
F_{ij} = \left.\sum_{X,Y}\sum_{\ell} \frac{\partial
C_{\ell}^{X}}{\partial\alpha_{i}}
({\bf{C}}^{XY}_{\ell})^{-1}\frac{\partial
C_{\ell}^{Y}}{\partial\alpha_{j}}\right|_{\bf{\alpha} =
\bar{\bf{\alpha}}}.
\end{equation}
We assume a gaussian beam profile and assume that foregrounds are
perfectly subtracted. The elements of the symmetric matrix
${\bf{C}}^{XY}_{\ell}$   are enumerated in \cite{Verde:2005ff}.
We specify a proposed experiment in terms of $\sigma_{b} =
\Theta_{\rm{FWHM}}/\sqrt{8\ln 2}$, the Gaussian beamwidth, where
$\Theta_{\rm{FWHM}}$ denotes the ``full width at half maximum''
power of the beam. The noise per multipole is $n_{0} =
\sigma^{2}_{pix}\Omega_{pix}$, where $\Omega_{pix}
=\Theta_{\rm{FWHM}}^{2} = 4\pi f_{sky}/N_{pix}$ is the beam solid
angle, $N_{pix}$ is the number of pixels (independent beams) in the
survey region and $f_{sky}$ is the fraction of the sky covered by
observations \cite{Verde:2005ff}.  The variance per pixel is
$\sigma_{pix}^{2}$ which can be obtained from the detector
sensitivity as $\sigma_{pix} = s/\sqrt{Nt}$, where $N$ is the number
of detectors and $t$ the integration time per pixel.

In this work we consider three cases:
\begin{enumerate}
\item A cosmic variance limited survey - $n_{0} = 0$, (C.V.),
\item The projected errors from the Planck satellite, (Plk),
\item The ideal satellite experiment of \cite{Verde:2005ff} (Ideal).
\end{enumerate}
For all the above experiments we take $\ell_{max}
= 1500$, unless stated otherwise.   For the Planck satellite, we follow
\cite{Eisenstein:1998hr} and \cite{Kinney:1998md} and use only two
of the ten available channels (143 and 217 GHz), assuming that the
remaining channels have been used for  foreground subtraction. This
simple estimate leads to somewhat optimistic predictions
for Planck's sensitivity to a primordial B-mode. However, since we
are interested in the {\em differences} between the standard
spectral analysis and slow roll reconstruction, it suffices for our
needs.  For the full sky ``Ideal'' experiments, we assume five
identical channels at frequencies 30, 50, 70, 100 and 200 GHz. The
satellite noise specifications are shown in Table \ref{sat}.
\begin{table}
\centering
\begin{tabular}{r| c c c c c}
\hline\hline
Experiment & $f_{sky}$ & Frequency & $\theta_{beam}$ & $\sigma_{T}$
& $\sigma_{P} $\\
& & (GHz) & ($'$) & $(\mu K)$ & $(\mu K)$
\\\hline
\multirow{2}{*}{Planck (Plk)} &  \multirow{2}{*}{0.65} & 143 & 8.0 & 5.2 & 10.8 \\
& & 217 & 5.5 & 11.7 & 24.3 \\
\hline
Ideal Sat (Ideal). & 0.8 & 30 - 200 & 8.0 & 2.2 & 2.2 \\
\hline\hline
\end{tabular}
\caption{Experimental specifications for CMB satellites.}\label{sat}
\end{table}
In all cases (except where explicitly specified) we include the
effects of gravitational lensing of the $E$-mode polarization into
$B$-mode polarization \cite{Zaldarriaga:1998ar}, which is computed
within \texttt{CAMB}. We do not consider cases in which the $B$-mode
has been delensed. We also tested our code by checking we could
recover the results of \cite{Verde:2005ff} in the cases which
overlap with our assumptions here.

\subsection*{Parameterizations}

The Fisher matrix formalism forecasts the likely error ellipse for
any given fiducial model -- and the size of this ellipse is a
function of the chosen central parameter values.    For
concreteness, we assume that the non-inflationary parameters are
well described by a $\Lambda$CDM cosmology.  In what follows we set
the central values of the baryon fraction, $\Omega_{b}$; the cold
dark matter fraction, $\Omega_{cdm}$; the reduced Hubble parameter,
$h$; and the optical depth to reionization, $\tau$; to their central
values found in the WMAP3 concordance cosmology.   As noted earlier,
consistency with an inflationary prior requires $\Omega_{\rm{Total}}
= 1$. Specifically we fix $\Omega_{\Lambda} =
1-\Omega_{cdm}-\Omega_{b}$ and assume the dark energy has the
equation of state $w=-1$. We ignore the neutrino mass.

We now compare models parameterized in terms of spectral variables,
(the amplitude, $A(k_{0})$; the scalar index, $n_s$; and the
running, $\alpha$; as well as the corresponding tensor variables) to
an HSR analysis.  We define our amplitude variable $A(k_{0})$ via
\begin{equation}
P(k_{0}) = 2.95\times 10^{-9} A(k_{0}).
\end{equation}
In the analysis below we always impose the simple
inflationary prior on the tensor spectrum  $n_{t} = -r/8$, where $r$
is the ratio of the tensor and scalar amplitudes.  This gives a
considerable advantage to the spectral variables, since a
generic treatment of the tensor modes would not assume this
correlation, whereas it is implicit in the HSR formalism. However,
it is simple to apply and used in most analyses of CMB data; so we
adopt it here.

The likely parameter constraints on spectral variables are well
understood (see \cite{Verde:2005ff} for instance), but this is the
first forecast of the expected errors for HSR parameters. We will
show that with very high quality data the HSR formalism results in a
more sharply peaked likelihood surface, as it can make use of higher
order correlations between slow roll variables. Since we can place
cuts on the HSR parameter space using an ``e-folds" prior, which is
not included in our Fisher matrix analysis, the following treatment
will necessarily underestimate the strength of the slow roll
reconstruction formalism. Ironically, this is particularly
noticeable with less precise data -- if the likelihood contours are
relatively large, they are more likely to extend into regions
excluded by an e-folds prior.

\subsection*{The Lever-arm effect}

It is intuitively clear that extending the range of wavenumbers
(typically measured in terms of the number of decades of $k$ spanned
by the data
) over which we have CMB data will tighten
the parameter bounds.  However, if the spectral indices are very
well approximated by the lowest order slow roll expressions ($n_s =
1+2\eta-4\epsilon$ and $r=16\epsilon$), and
$\epsilon$ and $\eta$ do not change significantly as observable
modes leave the horizon, the slow roll formalism
amounts to a linear transformation of the spectral variables.   In
terms of the spectral indices and running, the spectra take the form
\begin{equation}
P_{\mathcal{R}}(k) =
A^{2}_{s}\left(\frac{k}{k_{0}}\right)^{n_{s}(k)-1},
\end{equation}
where
\begin{equation}
n_{s}(k) =
n_{s}(k_{0})+\frac{\alpha}{2}\ln\left(\frac{k}{k_{0}}\right)+ \dots,
\end{equation}
and
\begin{equation}
P_{h}(k)  = A^{2}_{s}\;r \left(\frac{k}{k_{0}}\right)^{n_{t}},
\end{equation}
where $n_{t}$ is assumed to be scale invariant.

The extra information exploited by slow roll reconstruction comes
from the hierarchy of consistency relations that exist between the
spectral parameters at \emph{all} orders in slow roll. Consider, for
example, the case in which we have two spectral parameters,
$\{n_{s}, r\}$, which is analogous to the High-$\epsilon$,
2-Parameter inflationary model, specified via $\{\epsilon, \eta\}$.
In the spectral parametrization, $\alpha = 0$, but slow roll
requires
\begin{equation}\label{consist}
\alpha \approx
-\frac{1}{1-\epsilon}\left[8\epsilon^{2}-10\epsilon\eta\right]
\approx -\epsilon\left[8\epsilon-10\eta\right](1+\epsilon).
\end{equation}
The corresponding spectral model $\{n_{s},r\}$ has only the term
linear in $\ln(k/k_{0})$ in $\ln (P_{\mathcal{R}}(k))$, whereas the
slow roll parametrization has additional terms of the form
$\mathcal{O}(\epsilon, \eta, \ldots)\ln(k/k_{0})^{2}$.  Even though
the co-efficient is small, when the lever arm (in $\ln{k/k_0}$) is
large the quadratic terms will make a substantial contribution
to the spectrum.

One can show that, to a good approximation, derivatives of the
$C_\ell$ with respect to the parameters that specify the power
spectrum can be written as
\begin{equation}
\frac{1}{C_{\ell}}\frac{\partial C_{\ell}}{\partial \alpha_{i}}
\approx \left.\frac{1}{P(k)}\frac{\partial P(k)}{\partial
\alpha_{i}}\right|_{k = \ell/(\eta_{0}-\eta_{*})},
\end{equation}
where $\alpha_{i} = n_s, dn_{n}/d\ln k, \epsilon, \eta$ etc and
$P(k)$ is the appropriate spectrum (scalar for temperature and
$E$-mode, tensor for $B$-mode). Here the quantity
$\eta_{0}-\eta_{*}$ denotes the conformal time interval between
today ($\eta_{0}$) and last scattering ($\eta_{*}$) and is not to be
confused with the slow roll parameter $\eta$. The elements of
$F_{ij}$ for which $\alpha_i$ and $\alpha_j$ are both spectral
parameters are approximately
\begin{equation}
F_{ij} = \sum_{\ell =
2}^{\ell_{max}}\frac{(2\ell+1)}{2}\left.\frac{\partial \ln
P_{\mathcal{R}}(k)}{\partial \alpha_{i}}\frac{\partial \ln
P_{\mathcal{R}}(k)}{\partial \alpha_{j}}\right|_{k =
\ell/(\eta_{0}-\eta_{*})} \, .
\end{equation}
If we are working with $\{n_{s}, \alpha, r\}$  we have
\begin{eqnarray}
\frac{\partial\ln C^{T}_{\ell}}{\partial n_{s}} & \approx &
\frac{\partial\ln
P_{\mathcal{R}}(k)}{\partial n_{s}} = \ln\left(\frac{k}{k_{0}}\right),\\
\frac{\partial\ln C^{T}_{\ell}}{\partial \alpha} & \approx &
\frac{\partial\ln P_{\mathcal{R}}(k)}{\partial \alpha} =
\frac{1}{2}\ln\left(\frac{k}{k_{0}}\right)^{2},\\
\frac{\partial\ln C^{T}_{\ell}}{\partial r} & \approx & 0\quad,
\end{eqnarray}
where the last line signifies we are neglecting the contribution of
tensor modes to the {\em temperature\/} anisotropies. Assuming the
consistency relation,
\begin{eqnarray}
\frac{\partial\ln C^{B}_{\ell}}{\partial n_{s}} & \approx &
\frac{\partial\ln P_{h}(k)}{\partial n_{s}} = 0,\\
\frac{\partial\ln C^{B}_{\ell}}{\partial \alpha} & \approx &
\frac{\partial\ln P_{h}(k)}{\partial \alpha} =
0, \\
\frac{\partial\ln C^{B}_{\ell}}{\partial r} & \approx &
\frac{\partial\ln P_{h}(k)}{\partial r } = 1
-\frac{1}{8}\ln\left(\frac{k}{k_{0}}\right).
\end{eqnarray}
The corresponding slow roll parameter set is
$\{\epsilon,\eta,\xi\}$. Keeping the leading terms in slow roll
multiplying each power of $\ln{k/k_0}$ we find
\begin{eqnarray}\nonumber
\frac{\partial\ln C_{\ell}}{\partial \epsilon} & \approx &
\frac{\partial\ln P_{\mathcal{R}}(k)}{\partial \epsilon} = ( -
4-4(1+\mathcal{C})\epsilon-\frac{1}{2}(3-5\mathcal{C})\eta)\ln\left(\frac{k}{k_{0}}\right)\\
&
&+\frac{1}{2!}\left[10\eta-16\epsilon+\frac{1}{2}(5-7\mathcal{C})\xi\right]\ln\left(\frac{k}{k_{0}}\right)^{2}
-\frac{14}{3!}\xi\ln\left(\frac{k}{k_{0}}\right)^{3},\\\nonumber
\frac{\partial\ln C_{\ell}}{\partial \eta} & \approx &
\frac{\partial\ln P_{\mathcal{R}}(k)}{\partial \eta} =(2
-\frac{1}{2}(3-5\mathcal{C})\epsilon)\ln\left(\frac{k}{k_{0}}\right)\\&&+
\frac{1}{2!}\left(10\epsilon-\frac{1}{2}(3-\mathcal{C})\xi\right)\ln\left(\frac{k}{k_{0}}\right)^{2}
+\frac{2}{3!}\xi\ln\left(\frac{k}{k_{0}}\right)^{3}, \\\nonumber
\frac{\partial\ln C_{\ell}}{\partial \xi} & \approx &
\frac{\partial\ln P_{\mathcal{R}}(k)}{\partial \xi}
=\frac{1}{2}(3-\mathcal{C})\ln\left(\frac{k}{k_{0}}\right)-
\frac{1}{2!}\left(2-\frac{1}{2}(5-7\mathcal{C})\epsilon\right.\\
&&\left.+\frac{1}{2}(3-\mathcal{C})\eta\right)
\ln\left(\frac{k}{k_{0}}\right)^{2}
+\frac{1}{3!}(2\eta-14\epsilon+(3-\mathcal{C})\xi)\ln\left(\frac{k}{k_{0}}\right)^{3},
\end{eqnarray}
where in all instances, $k$ is understood to be $k =
\ell/(\eta_{0}-\eta_{*})$. Similar expressions exist for the tensor
spectrum.  As noted above, for a given number of parameters, the
slow roll parametrization  adds an extra power of $\ln(k/k_{0})$,
which is an expression of a higher order consistency condition
\cite{Copeland:1993zn,Kosowsky:1995aa,Lidsey:1995np}. This extra power provides more leverage
on the likelihood space as one adds information further from the
fiducial scale.

To compare the constraining power of the two parameterizations, we
consider the volume of parameter space contained within the Fisher
ellipses, as a function of $\ell_{max}$, in analogy to the metric
developed by the Dark Energy Task Force  \cite{Albrecht:2006um}. For
three slow roll or spectral parameters, this is the volume of an
ellipsoid.  In order to make a comparison between the two different
parameterizations, we normalize the volume by its value when
$\ell_{max}$ is set to the pivot scale which, for $k_{0} = 0.05$
Mpc$^{-1}$, is $\ell\sim500$. The constraining power of the
parametrization determined by how quickly the volume contracts as
information is added, and we plot a Figure of Merit, defined as the
inverse of the normalized $n$-dimensional volume enclosed within the
error ellipsoids.
%
%
\begin{figure*}
\begin{center}
\resizebox{12cm}{!}{\includegraphics{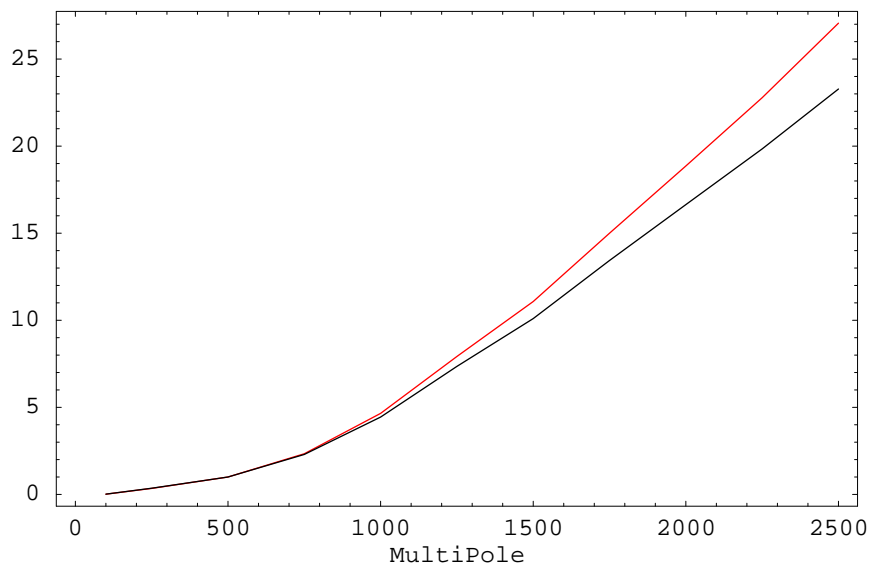}}\\
\end{center}
\caption{Model with $\epsilon= 0.015$, $\eta = -0.02$, $\xi = 0.0$.
The figures of merit for the $\{\epsilon, \eta,
\xi \}$ (red) and $\{r, n_{s}, \alpha\}$ parameterizations (black),
normalized at $\ell = 500$, calculated using the full Fisher matrix
with cosmic variance only.  As the lever arm in $\ell$ becomes
longer, the extra information in the slow roll formalism leads to a
more sharply peaked likelihood function and thus tighter constraints
on the model parameters. }\label{extreme05}
\end{figure*}
%
%
In reality, the CMB probes a relatively small range of scales (the
largest value of $\ln{(k/k_0)}$ is likely on the order of 6), which
limits the maximal extent of our lever arm.  Moreover, the highest
order term in  $\ln{(k/k_0)}$ in the derivatives of the $C_l$ is
multiplied by a slow roll parameter, which is necessarily
considerably smaller than unity.  Consequently, if we restrict
attention {\em solely\/} to CMB data, we will need a very accurate
measurement of the primordial CMB to make use of the extra
information. To examine our hypothesis, however, we can look at
extreme cases. From equation (\ref{consist}), the largest
deviations are  found in models where $\epsilon$ is large, and
$\epsilon$ and $\eta$ have opposite signs. For example, consider
$\epsilon = 0.015$ and $\eta = -0.02$, corresponding to spectral
parameters $r = 0.23$, $n_{s}= 0.9$ and $\alpha = 0.0$. Figure
\ref{extreme05} shows the Figures of Merit for the spectral and slow
roll formulations for this parameter set, and the two values have
begun to diverge  once $\ell_{max} > 1500$.  This is a gedanken
experiment, given both that we assumed foreground subtraction and
that $n_s-1$ and $r$ are unrealistically large. This illustrates
that second order consistency conditions are probably not testable
by CMB data alone. However, it also gives cause for optimism that
this lever arm effect can be exploited by combinations of   CMB
surveys and datasets sensitive to the primordial spectrum at shorter
scales.

Adding further decades in information from the power spectrum will
increase the difference between the figures or merit.  However,  we
also see that the running induced solely by $\epsilon$ and $\eta$
being non-zero is proportional to $\epsilon$. Thus in Low-$\epsilon$
models (corresponding to those with a negligible value of $r$),
increasing the lever arm in $\ln{k/k_0}$ will not give any {\em
extra} advantage to slow roll reconstruction, relative to the
spectral variables, unless we are lucky enough to see scale dependence $\xi$ induced by a non-zero $\eta$. Moreover, if we do observe a running in conjunction with a low value of $r$, then we will obtain an unambiguous measurement of $\xi$.

\section{Forecasts}\label{results}

We use the derivative methodology outlined in
\cite{Eisenstein:1998hr} to compute the numerical derivatives using
\textrm{CAMB} \cite{Lewis:1999bs}. We consider the parameters
`class-by-class' as described in \S \ref{classificationsection},
beginning with the simplest possible case, the High-$\epsilon$
1-Parameter model.  In each case where one or more of the parameters
is zero, we consider both situations in which we fit for the
parameter at its zero value and the alternative in which it is not
included in the parameter set. This second case reduces the
dimensionality of the parameter space by one and naturally leads to
tighter restrictions on the other parameters. For all but the
High-$\epsilon$ 1-Parameter  models, we work with a fiducial point
chosen so that  $n_{s} = 0.97$ at the pivot. That is, $r$ sets
$\epsilon$ while $\alpha$ sets $\xi$ and then both $\epsilon$ and
$\xi$ are used to set $\eta$ keeping $n_{s} = 0.97$.

\begin{table*}
\begin{tabular}{|c|l|llllll|}\hline
Parameter & Fiducial & C.V. Abs  & C.V. \% & Ideal Abs. & Ideal \% & Plk Abs & Plk \% \\\hline
$\omega_{b}$ & 0.024& 0.0000607 & 0.253 &  0.0000772 & 0.322 & 0.00018 &  0.75  \\
$\omega_{cdm}$ & 0.12 & 0.000293 & 0.244 & 0.000503 & 0.419 & 0.0014 & 1.17  \\
$H_{0}$ & 72.0 & 0.132 &  0.184 & 0.225 & 0.312 & 0.664 & 0.922 \\
$\tau$ & 0.164& 0.0026 & 1.59 & 0.00306 & 1.87  & 0.000637 & 3.88 \\
$A(k_{0})$ & 0.9 &  0.00437 & 0.486 & 0.00547 & 0.607 &  0.0116 & 1.19
\\\hline\hline
$\epsilon$ & 0.01 & 0.000122 & 1.22 &0.00021 & 2.1 & 0.00181 & 18.1 \\
$\eta$ &  0.0 & 0.00151 & - & 0.00206 &-  & 0.00331 & -  \\
$\xi$ & 0.0 & 0.00174 & - & 0.00216 &- & 0.00285 & - \\\hline\hline
$ \epsilon $ & 0.01 &  0.000122 & 1.22 & 0.000209 & 2.09 & 0.00179 & 17.9 \\
$\eta $ & 0.0 & 0.000871 & - & 0.00115 & - & 0.00417 & - \\\hline\hline
$\epsilon$ & 0.01 &  0.000116 &  1.16 & 0.000193 & 1.93 &  0.000914 & 9.14 \\\hline
\end{tabular}
\caption{Errors for model: $\epsilon = 0.01$, $\eta = 0.0$, $\xi =
0.0$. This model gives $49.5$ e-folds of inflation after the
fiducial scale leaves the horizon.  Here C.V. refers to a cosmic
variance limited survey, Ideal is the ``straw man'' satellite
proposal of \cite{Verde:2005ff} and Plk refers to Planck (although
this analysis assumes perfect foreground subtraction, and thus
over-estimates Planck's capabilities, especially at small $r$).  We
give absolute and percentage errors for each parameter. In all cases
we assume $\ell_{max} = 1500$.    The second block gives the
forecast for a fit to $\{\epsilon,\eta,\xi\}$, the middle block for
a fit to $\{\epsilon,\eta\}$ and the bottom block give the results
for a fit to $\epsilon$ alone. As expected the constraints get
tighter as the number of parameters is reduced. Finally the top
block gives the forecasts for the other cosmological parameters,
derived from the three parameter slow roll fit. \label{eps01} }
\end{table*}

\begin{table*}
\begin{tabular}{|c|l|llllll|}\hline
Parameter & Fiducial & C.V. Abs  & C.V. \% & Ideal Abs. & Ideal \% & Plk Abs & Plk \% \\\hline
$\epsilon$ & 0.0025 & 0.0000522 & 2.09 & 0.000112 & 4.48 &  0.000826 & 33.\\
$\eta$ &  0.0 & 0.00151 & -  & 0.00206 &  -  &  0.00369 & - \\
$\xi$ & 0.0 &  0.00172 &  - & 0.00214 & - & 0.00332 & -  \\\hline\hline
$ \epsilon $ & 0.0025 & 0.0000522 & 2.09  &  0.000112 & 4.48 & 0.000825 & 33 \\
$\eta $ & 0.0 & 0.00083 &  -  &  0.00108 & -  & 0.00266 & - \\\hline\hline
$\epsilon$ & 0.0025 & 0.0000517 &  2.07 &  0.000109 & 4.38 & 0.000665 & 26.6 \\\hline
\end{tabular}
\caption{Errors for model: $\epsilon = 0.0025$, $\eta = 0.0$, $\xi =
0.0$. This model gives rise to $199.5$ e-folds of inflation after the
fiducial scale leaves the horizon, and we have used the same
conventions as Table~\ref{eps01}. \label{eps0025}   }
\end{table*}

\subsubsection*{High-$\epsilon$ 1-Parameter  Models:}

 This is perhaps the simplest  possible model of inflation
and effectively describes an $m^{2}\phi^{2}$ potential. However, the
actual portion of the potential sampled by inflation is not
specified, and if this turns out to be far from the minimum
inflation must end via a hybrid transition.  We consider $\epsilon =
0.01$ and $\epsilon = 0.0025$.  Via equation \ref{Nendeps} the first
case has  a further 49.5 e-folds of inflation after the fiducial
scale has left the horizon and gives a tilt of $n_{s} = 0.96$. The
second case has 199.5 further e-folds and a tilt of $n_{s} = 0.99$.
This model has no obvious analogue in spectral variables, so we only
present forecasts for the slow roll  variables.  As can be seen from
Tables \ref{eps01} and \ref{eps0025}, this one parameter model can
be very tightly constrained. However, the precise measurement of $r$
is facilitated by the correlation of $r$ and $n_s$,
and the latter is very tightly constrained by measurements of the
temperature anisotropies.

\subsubsection*{Low-$\epsilon$ 1-Parameter  Models:}
In this case, the tensor signal is tiny, and $|\eta| \ne 0$.
Table~\ref{epsem10etam015B} shows the forecast errors  for a model
$\eta$ so that $n_{s} = 0.97$ and $\alpha = 0$, so $\eta = -0.015$
and $\xi= 0.0$.  This scenario model leads to $\sim 600$ e-folds of
inflation, so we would  clearly require  a hybrid transition. Note
that when $\xi$ and $\alpha$ are marginalized over, the relative
constraints on $n_{s}-1$ and $\eta$ are identical. We make no
forecast for $\epsilon$ -- this fit is analogous to that of a pure
$\Lambda$CDM cosmology. This forecast includes the $B$-mode
contribution from the lensed $E$-mode.  We present the same forecast
in Table~\ref{epsem10etam015}, with the $B$-mode spectrum omitted
from our analysis  -- that is all derivatives of $C^{BB}$ are
assumed to vanish.   We see that in the absence of primordial
tensors, a highly accurate $B$-mode measurement would significantly
improve constraints on $\omega_{cdm}$, since the lensing is induced
by the dark matter potential wells  -- although whether this level
of accuracy is achievable in practice is of course a very different
matter.

We give forecasts for both slow roll and spectral variables. We
construct the analogous spectral variables by computing them at the
pivot. The resulting spectra are subtly different, since a two
parameter slow roll model has a non-zero running if $\epsilon$ is
not minuscule; this effect will necessarily be absent in the
spectral case.  If we assume that the underlying cosmology is  a
Low-$\epsilon$ 1-Parameter model,  we are still free to include
$\xi$ and $\alpha$ in our parameter set.  In this case, $\alpha
\approx -2\xi$ and we see that the constraints on $\xi$ indeed match
those on $\alpha$. However, adding $\xi$ {\em weakens\/} the
constrain on $\eta$ since $n_s$ has a partial degeneracy between
$\alpha$ and $\eta$. This amounts to a rotation in our parameter
space, and does not imply that slow roll constraints are
intrinsically less accurate than the spectral constraints. We
implement ``Low-$\epsilon$'' scenarios by using the full slow-roll
hierarchy with   $\epsilon$ set to a very low value  ($10^{-10}$).
The results are insensitive to  the actual value of $\epsilon$,
provided it is far below the threshold of detectability.

\begin{table*}[p]
\begin{tabular}{|c|l|llllll|}\hline
Parameter & Fiducial & C.V. Abs  & C.V. \% & Ideal Abs. & Ideal \% &
Plk Abs & Plk \% \\\hline
$\omega_{b}$ & 0.024 & 0.0000614 & 0.256 & 0.0000778 & 0.324 & 0.000182 & 0.758 \\
$\omega_{cdm}$ & 0.12 & 0.000311 & 0.259 & 0.00051 & 0.425 & 0.00142 & 1.19 \\
$H_{0}$ & 72. & 0.141 & 0.195 & 0.229 & 0.317 & 0.676 & 0.939 \\
$\tau$ & 0.164 & 0.00281 & 1.71 & 0.00328 & 2. & 0.00647 & 3.94 \\
$A(k_{0})$ & 0.9 & 0.00468 & 0.52 & 0.00577 & 0.641 & 0.0117 & 1.3
\\\hline\hline
$n_{s}-1$ & -0.03 & 0.00166 & 5.53 & 0.00215 & 7.16 & 0.00439 &  14.3\\
$\alpha$ & 0. & 0.00344 & -  & 0.00428 & -  & 0.0066 & -
\\\hline\hline
$n_{s}-1$ & -0.03 & 0.00166 & 5.53 & 0.00212 & 7.07 & 0.00427 & 14.2\\\hline\hline
$\eta$ & -0.015 & 0.00308 & 20.6 & 0.00404 & 26.9 & 0.00632 & 42.1 \\
$\xi$ & 0. & 0.00346 & -  & 0.00432 & -  & 0.00669 & -  \\\hline\hline
$\eta$ & -0.015 & 0.000828 & 5.52 & 0.00106 & 7.05 & 0.00213 & 14.2 \\\hline
\end{tabular}
\caption{Errors for model $\epsilon = 10^{-10}$, $\eta = -0.015$ and
$\xi = 0.0$, where the CMB spectra include the $B$-mode contribute
from the lensed $E$ mode. The conventions here match those of
Table~\ref{eps01}, although we also give forecasts for the
corresponding spectral parameters. \label{epsem10etam015B}}
\end{table*}

\begin{table*}
\begin{tabular}{|c|l|llllll|}\hline
Parameter & Fiducial & C.V. Abs  & C.V. \% & Ideal Abs. & Ideal \% &
Plk Abs & Plk \% \\\hline
$\omega_{b}$ & 0.024 & 0.000066 & 0.275 & 0.0000831 & 0.346 & 0.000192 & 0.801 \\
$\omega_{cdm}$ & 0.12 & 0.000634 & 0.528 & 0.000755 & 0.63 & 0.00155 & 1.29 \\
$H_{0}$ & 72. & 0.271 & 0.376 & 0.33 & 0.459 & 0.741 & 1.03 \\
$\tau$ & 0.164 & 0.00297 & 1.81 & 0.00338 & 2.06 & 0.00642 & 3.92 \\
$A(k_{0})$ & 0.9 & 0.00574 & 0.638 & 0.00657 & 0.73 & 0.00117 & 1.3\\\hline
$n_{s}-1$ & -0.03 & 0.00204 & 6.8  & 0.00248 & 8.23 & 0.00459 & 15.3 \\
$\alpha$ & 0. & 0.00377 & -  & 0.00443 & -  & 0.00653 & -
\\\hline\hline
$n_{s}-1$ & -0.03 & 0.00198 & 6.6 & 0.00237 & 7.9 & 0.00457 & 15.2 \\\hline\hline
$\eta$ & -0.015 & 0.00363 & 24.2 & 0.00436 & 29.1 & 0.00632 & 42.1 \\
$\xi$& 0. & 0.0038 & -  & 0.00448 & -  & 0.00662 & -  \\\hline\hline
$\eta$ & -0.015 & 0.00099 & 6.6 & 0.00119 & 7.9 & 0.00228 & 15.2 \\\hline
\end{tabular}
\caption{Errors for model $\epsilon = 10^{-10}$, $\eta = -0.015$ and
$\xi = 0.0$. In this case we did not include any $B$-mode
information in our forecasts. The forecasts for our measurement of
$\omega_{cdm}$ are considerably looser than those found in
Table~\ref{epsem10etam015B} even though the primordial $B$ mode is
entirely absent, since the amplitude of the $B$ mode is now entirely
dependent on the depth of the dark matter potential wells, and we
are not exploiting this information here.}\label{epsem10etam015}
\end{table*}

\subsubsection*{High-$\epsilon$ 2-Parameter  Models:}

If we consider only scalar perturbations  High-$\epsilon$
2-Parameter models  have a degeneracy in the  $\epsilon-\eta$ plane,
which is broken by the limits on $\epsilon$ derived from $B$-mode
observations, and/or the running of the spectral index.  As we
showed in Figure \ref{extreme05}, the constraints on the slow roll
parameters will tighten more rapidly than the analogous ones on the
spectral parameters when $\epsilon$ is relatively large thanks to
the lever arm effect.  In the forecast presented in
Table~\ref{eps015eta015xi0} we  have set $\epsilon = 0.015$ and
chosen $\eta$ so that $n_s = 0.97$.  However, since our forecasts
have been done with $\ell_{max} = 1500$ we do not see the benefit of
the lever arm effect here.

\begin{table*}
\begin{tabular}{|c|l|llllll|}\hline
Parameter & Fiducial & C.V. Abs  & C.V. \% & Ideal Abs. & Ideal \% & Plk Abs & Plk \% \\\hline
$\omega_{b}$ & 0.024& 0.0000607 & 0.254 &  0.0000772 & 0.322 & 0.00018 &  0.75  \\
$\omega_{cdm}$ & 0.12 & 0.000305 & 0.254 & 0.000516 & 0.43 & 0.00141 & 1.18 \\
$H_{0}$ & 72.0 & 0.136 &  0.189 & 0.23 & 0.319 & 0.67 & 0.93 \\
$\tau$ & 0.164& 0.00271 & 1.66 & 0.00318 & 1.94  & 0.0064 & 3.9 \\
$A(k_{0})$ & 0.9 &  0.00451 & 0.501 & 0.00558 & 0.62 &  0.0116 &  1.29 \\\hline\hline
$ r $ & 0.24 & 0.0026 & 1.08 & 0.00425 & 1.77 & 0.351 & 14.6 \\
$ n_{s}-1 $ & -0.03 & 0.00167 & 5.57 & 0.00217 & 7.23 & 0.00427 & 14.2 \\
$\alpha $ & 0.0 & 0.00355 & - & 0.0044 & - & 0.00675 & - \\\hline\hline
$r $ & 0.24 & 0.0026 & 1.08 & 0.00425 & 1.77 & 0.0348 & 14.5\\
$n_{s}-1$ & -0.03 & 0.00167 & 5.57 & 0.00213 & 7.1 & 0.00425 & 14.2
\\\hline\hline
$\epsilon$ & 0.015 & 0.000158 & 1.06 & 0.000259 & 1.73 & 0.00213 & 14.3 \\
$\eta$ &  0.015 & 0.00154 & 10.3 & 0.00209 & 13.9  & 0.00425 & 34.1  \\
$\xi$ & 0.0 & 0.00176 & - & 0.00218 &- & 0.00337 & - \\\hline\hline
$ \epsilon $ &0.015 & 0.000157 & 1.05 & 0.000257 & 1.72 & 0.00269 & 13.9 \\
$\eta $ &  0.015 & 0.000911 & 6.08 & 0.00122 & 8.11  & 0.00483 & 32.2  \\\hline
\end{tabular}
\caption{Errors for model: $\epsilon = 0.015$, $\eta = 0.015$, $\xi
= 0.0$}\label{eps015eta015xi0}
\end{table*}

\subsubsection*{Low-$\epsilon$ 2-Parameter Models:}

We now consider the scenario in which the inflaton potential is specified by $\eta$ and
$\xi$. That is, $r=0$, $n_{s}\neq 1$ and $\alpha \neq 0$. In this
case, the slow roll parametrization  is (almost) identical to the
spectral parametrization. There will be terms proportional to $\xi
\ln(k/k_{0})^{3}$, but these are never large over the range of
scales probed by the CMB.  In computing the errors for these models,
we do not include information from lensing of the $E$-mode
polarization into $B$-mode polarization, so we have somewhat
over-estimated the likely errors for $\omega_{cdm}$. However, both
the spectral and slow roll variables will be equally affected.

The parameter $\xi$ dominates the dynamics very quickly when $\xi>0$
\cite{Chongchitnan:2005pf}, and thus $\alpha<0$. In
\ref{lowscale} we present an analytic treatment of the slow roll
dynamics for the Low-$\epsilon$ 2-Parameter  model, from which we
can compute the remaining number of e-folds. Fixing $n_{s} = 0.97$
and then decreasing $\alpha$ from zero leads to a dramatic decrease
in $N_e$, as shown in Figure \ref{plot2}.  For $\eta = -0.015$, $\xi
= 0.0$ inflation ends after $\sim600$ e-folds, while $\xi = 10^{-5}$
gives $\sim 260$ e-folds and $\xi = 0.001$ yields only $\sim 55$
efolds, even though the corresponding value of $\alpha$ at the pivot
is very small.   At this point we are beginning to bite into the
e-foldings window of equation (\ref{Nefolds}). Almost any prior on
the number of e-folds will lead to tight constraints on $\xi$
\cite{Easther:2006tv}. This constraint has has no analog when the
power spectrum is specified in terms of $n_s$, and $\alpha$, and
allows slow roll reconstruction to impose much tighter constraints
than fits to $n_s$, and $\alpha$.

For this set of parameters, we find that the smallest detectable
value of $\xi$ (assuming a cosmic variance limited measurement of
the CMB)  is $\xi \sim 0.002$. Consequently, if $\xi$ can be
detected in via fits to CMB data, the end of inflation was `in
sight' as the fiducial modes left the horizon, and we do not need to
posit a hybrid transition. Given the difficulties of foreground
subtraction, a constraint this tight would almost certainly require
a combined fit to several orthogonal datasets. Furthermore, as
pointed out by \cite{Hamann:2006pf}, the extra parameters we have
ignored (neutrino mass, dark matter equation of state), can have a
non-negligible effect on estimates of primordial spectrum parameters
and thus forecasts. We demonstrate our ability to constrain $\xi$ in
Tables \ref{epsem10etam016xi002}-\ref{epsem10etam037xi03}. In each
case we forecast the errors with $\eta$ and $\xi$ chosen so that
$n_{s} = 0.97$, and with increasing values of $\xi$.

\begin{table*}
\begin{tabular}{|c|l|llllll|}\hline
Parameter & Fiducial & C.V. Abs  & C.V. \% & Ideal Abs. & Ideal \% & Plk Abs & Plk \% \\\hline
$n_{s}-1$ & -0.03 & 0.002 & 6.67 & 0.00247 & 8.23  & 0.00459 & 15.3  \\
$\alpha$ &-0.004 & 0.00378 & 94.4 & 0.0044 & 111. & 0.00653 & 163. \\\hline\hline
$\eta$ & -0.0165 & 0.0019 & 11.6 & 0.00228 & 13.8 & 0.00343 & 20.9 \\
$\xi$ & 0.002 & 0.0019 & 97.1 & 0.00225 & 112. & 0.00331 & 165. \\\hline
\end{tabular}
\caption{Errors for model: $\epsilon = 10^{-10}$, $\eta = -0.0165$,
$\xi = 0.002$}\label{epsem10etam016xi002}
\end{table*}

\begin{table*}
\begin{tabular}{|c|l|llllll|}\hline
Parameter & Fiducial & C.V. Abs  & C.V. \% & Ideal Abs. & Ideal \% & Plk Abs & Plk \% \\\hline
$n_{s}-1$ & -0.03 & 0.002 & 6.67 & 0.00236 & 7.87 & 0.00459 & 15.3 \\
$\alpha $ & -0.01 & 0.00378 & 37.8 & 0.00444 & 44.4 & 0.00655 & 65.5 \\\hline\hline
$\eta$ &  -0.019 & 0.0019 & 10.2 & 0.00228 & 12.2  & 0.00344 & 18.5 \\
$\xi$ & 0.005 & 0.00195 & 39 & 0.00225 & 45 & 0.00332 & 66.4 \\\hline
\end{tabular}
\caption{Errors for model: $\epsilon = 10^{-10}$, $\eta = -0.019$,
$\xi = 0.005$}\label{epsem10etam019xi005}
\end{table*}

\begin{table*}
\begin{tabular}{|c|l|llllll|}\hline
Parameter & Fiducial & C.V. Abs  & C.V. \% & Ideal Abs. & Ideal \% & Plk Abs & Plk \% \\\hline
$ n_{s}-1 $ & -0.03 & 0.002 &  6.67 & 0.00244 & 8.13 & 0.00457 & 15.2\\\hline
$\alpha $ & -0.06 & 0.00381 & 6.36 & 0 .00449 & 7.48 & 0.00672 & 11.2 \\\hline\hline
$\eta$ &  -0.037 & 0.00182 & 4.91 & 0.00222 & 6.02  & 0.00335 & 9.09 \\\hline
$\xi$ & 0.03 & 0.00189 & 6.31 & 0.00223 & 7.44 & 0.00333 & 11.1 \\\hline
\end{tabular}
\caption{Errors for model: $\epsilon = 10^{-10}$, $\eta = -0.037$,
$\xi = 0.03$}\label{epsem10etam037xi03}
\end{table*}

\subsubsection*{High-$\epsilon$ 3-Parameter  Models:}

Finally, we consider the case in which the first three slow roll
parameters are non-zero. In this case $\epsilon$ can be large, and
the slow roll variables can again evolve within the observable
window.  We present results for two representative cases in
Tables \ref{eps015eta014xi001} and \ref{eps015eta0075xi01}. However,
the principal constraining power of slow roll reconstruction,
relative to the usual spectral formulation, is again its ability to
include restrictions based on the duration of inflation. For
example, the model in Table \ref{eps015eta0075xi01} would manage a
bare 12 e-folds of inflation. If we {\em found\/} these central values for
$\{\epsilon,\eta,\xi\}$, we would know this description of the
inflaton potential was either incomplete or that the primordial
spectrum was not generated by slow
roll inflation.

\begin{table*}
\begin{tabular}{|c|l|llllll|}\hline
Parameter & Fiducial & C.V. Abs  & C.V. \% & Ideal Abs. & Ideal \% & Plk Abs & Plk \% \\\hline
$ r $ & 0.24 & 0.0026 & 1.08 & 0.00424 & 1.77 & 0.0351 & 14.6 \\\hline
$ n_{s}-1 $ & -0.03 & 0.00167 & 5.57 & 0.00216 & 7.2 & 0.00427 & 14.2\\\hline
$\alpha $ & -0.002 & 0.00349 & 175 & 0 .00436 & 218 & 0.00669 & 335 \\\hline\hline
$\epsilon$ & 0.015 & 0.000158 & 1.05 & 0.000258 & 1.72 & 0.00196 & 13.1 \\\hline
$\eta$ &  0.14 & 0.00154 & 11.0 & 0.00209 & 14.9  & 0.00489 & 34.9 \\\hline
$\xi$ & 0.001 & 0.00176 & 176 & 0.00218 & 218 & 0.00336 & 336 \\\hline
\end{tabular}
\caption{Errors for model: $\epsilon = 0.015$, $\eta = 0.014$, $\xi
= 0.001$. This model gives $\sim 33$ e-foldings of inflation after
the fiducial scale leaves the horizon.}\label{eps015eta014xi001}
\end{table*}

\begin{table*}
\begin{tabular}{|c|l|llllll|}\hline
Parameter & Fiducial & C.V. Abs  & C.V. \% & Ideal Abs. & Ideal \% & Plk Abs & Plk \% \\\hline
$ r $ & 0.24 & 0.00257 & 1.08 & 0.00422 & 1.78 & 0.0348 & 14.6 \\\hline
$ n_{s}-1 $ & -0.03 & 0.00167 & 5.57 & 0.00216 & 7.2 & 0.00427 & 14.2\\\hline
$\alpha $ & -0.02 & 0.00357 & 17.9 & 0 .00442 & 22.1 & 0.00683 & 34.2 \\\hline\hline
$\epsilon$ & 0.015 & 0.000158 & 1.06 & 0.00026 & 1.73 & 0.00213 & 14.2 \\\hline
$\eta$ &  0.0075 & 0.00152 & 20.4 & 0.00207 & 27.8  & 0.00509 & 68.3 \\\hline
$\xi$ & 0.01 & 0.00176 & 17.6 & 0.00218 & 21.8 & 0.00337 & 33.7 \\\hline
\end{tabular}
\caption{Errors for model: $\epsilon = 0.015$, $\eta = 0.0075$, $\xi
= 0.01$. This model gives only $\sim12$ e-foldings of inflation
after the fiducial scale leaves the
horizon.}\label{eps015eta0075xi01}
\end{table*}

\section{Conclusions}\label{concl}

We have considered the constraints one can place on the inflationary
parameter space via slow roll reconstruction.  In the course of this
work we also analyzed the precision of different approximation
schemes that can be used to compute the underlying spectrum. As
noted in \cite{Peiris:2006ug}, the primordial spectrum is easily
computed by numerically solving the mode equations
(\ref{uk}-\ref{vk}). In this case slow roll reconstruction contains
no {\em approximations\/}, as the slow roll hierarchy itself is
solved analytically \cite{Liddle:2003py}. For
almost all models of interest, approximate solutions will be more
than adequate given the precision of current data. The principal
concern is that different schemes for computing the spectrum will
impose implicit priors on the duration of inflation, and {\em
these\/} will effect the computed likelihood contours if they are
not properly understood. This issue was investigated in Refs.
\cite{Lesgourgues:2007aa,Hamann:2008pb}, which implement slow roll
reconstruction with an exact solver for the perturbation spectrum;
they find that the sole difference in likelihood contours using
current data can be attributed to implicit priors on the {\em
duration\/} of inflation. However, since the spectrum can be
computed numerically at little cost, using the exact solver has the
advantage of removing a possible source of uncertainty.

In addition, we introduce a scheme for classifying inflationary
models, based on the order at which the slow roll hierarchy is
truncated, and whether the $\epsilon$ parameter is small enough for
its contribution to the inflationary dynamics to be ignored. We
refer to these models as ``Low-$\epsilon$, N-parameter'' and
``High-$\epsilon$, N-parameter'' models. In the former case, we can
impose the slow roll prior while ignoring a possible tensor
contribution to the CMB, in  analogy with the usual $\Lambda$CDM
parameter set of the concordance cosmology.   Physically, $\epsilon$
responds to the ratio of the amplitude of tensor and scalar
perturbations, which reflects the energy scale at which inflation
occurs.  It is possible (and some might say likely, if one expects
inflation to be consistent with the microphysical constraints of
string theory) the inflationary scale is well below the GUT scale. In this case  $\epsilon$ and the amplitude of the tensor spectrum must be
vanishingly small. In this limit $\epsilon$
effectively decouples from the slow roll hierarchy.

Slow roll reconstruction can lead to tighter constraints on the
parameter space than those found with the usual spectral indices and
runnings. In particular, slow roll reconstruction allows us to
impose constraints on the total duration of inflation, because slow
roll reconstruction {\em requires\/} one to compute the duration of
the inflationary era for a given set of parameters. If this value is
very low -- as it can be for apparently reasonable parameters -- we
do not have a self-consistent description of inflation. This effect
is clearly seen in implementations of slow roll reconstruction
\cite{Peiris:2006sj,Peiris:2006ug}. Constraints based on the
duration of inflation imply a sharp cut on the relevant parameter
space, and this effect is not captured by our Fisher matrix
forecasts which assume a Gaussian likelihood. These constraints
simply encode our assumption that we are using enough slow-roll
parameters to provide a full description of the inflationary
potential, and do not introduce any  significant new or {\em ad hoc}
assumptions into the parameter estimates. The impact of this
constraint could be forecast by running Markov Chains on simulated
data for a proposed  experiment. We sketch an improved
``$\epsilon$-dependent" algorithm for imposing the ``e-fold prior''
that will be explored in detail in future work.

Slow roll reconstruction can also lead to improved constraints via
the lever-arm effect, which arises when the slow roll variables
evolve appreciably in the range of scales open to cosmological
experimentation. Higher order slow roll consistency relations are
automatically incorporated into slow roll reconstruction, and the
significance of these terms grows rapidly as once increases the
range of wavelengths over which we have information about the
primordial spectrum. Our error forecasts for the slow roll
parameters show that it is very unlikely this effect can be
exploited by datasets containing only CMB information.  However,
high redshift 21 cm data may probe the the primordial spectrum at
scales unreachable via the CMB. We thus conjecture that the
combination of CMB and 21 cm data may test these consistency
relations directly, fully realizing the power of slow roll
reconstruction. This conjecture is supported by the recent analysis
of \cite{Mao:2008ug}, who project that 21 cm data could provide a
4$\sigma$ detection of the intrinsic running predicted by simple
inflationary models. This is the same running one expects from
higher-order consistency conditions, which is the basis of the lever
arm effect we described earlier. There is thus good cause for
optimism that slow roll reconstruction will be able to exploit this
information to significantly tighten constraints on the inflationary
parameter space and we plan to explore this issue in detail in a
future publication. If inflation occurs well below the GUT scale,
$\epsilon$ is very small, and the tensor modes are likely to be
unobservable. The running in the spectral index induced by the scale
dependence of the slow roll parameters is {\em also\/} proportional
to $\epsilon$. Consequently, this effect is strongly correlated with
the inflationary scale. Finally, BBO style direct detection
experiments for gravitational waves probe the primordial spectrum at
wavelengths some 30 e-folds smaller than those in the CMB
\cite{Smith:2006xf} -- and in this case the running of the spectral
indices will be unmistakable if the tensor amplitude is large enough
to be seen, and slow roll reconstruction will be well-placed to take
advantage of these effects.

If the tensor to scalar ratio is vanishingly small, the tilt of the
scalar spectrum is fixed by $\eta$, if $|\xi| \ll |\eta|$. If $\eta =
-0.015$, corresponding to $n_{s} = 0.97$, inflation lasts for much
longer than 60 e-foldings. If our chosen set of variables provides a
complete description of the spectrum, inflation must end
via an some abrupt change in the potential, which cannot
be encoded in the truncated slow roll expansion.  Conversely, in
models where $\epsilon$ is non-trivial, or the $\xi$ parameter is
even slightly greater than zero, inflation will typically end within
60 e-folds, and often less.  In this case, the inflaton can `see'
the end of inflation well before it occurs.  Further, if $\epsilon$ is
constrained to be very small by a tight upper bound on $r$, any
running detected in the spectrum {\em must\/} be induced by an $\xi$
parameter, which is equivalent to learning that $V'''(\phi)$ is
non-trivial.

In summary, this paper provides explores the theoretical basis of
slow roll reconstruction, as originated by Easther and Peiris, and
extends our understanding of the properties of the slow roll
hierarchy.  Slow roll reconstruction requires no explicit
approximations in the treatment of the inflationary dynamics  and
exploits all possible sources of astrophysical information.
Consequently it provides an optimal solution to the problem of
reconstructing the inflationary potential from astrophysical data.
Whether the universe will be kind enough to permit this program to
be implemented in practice -- and whether the generation of the
primordial spectrum can in fact be attributed to an inflationary
model which can be described in terms of a single, slowly-rolling
field -- remains to be seen.

\ack We thank Simeon Bird, Jan Hamann, Will Kinney, Julien
Lesgourgues, Jonathan Pritchard,Wessel Valkenburg and Licia Verde for useful
discussions, and acknowledge the use of the spectrum code of
\cite{Lesgourgues:2007aa}.   We are particularly indebted to Hiranya
Peiris for providing the Hubble Slow Roll code, and for numerous
discussions as this paper was written.  PA and RE are supported in
part by the United States Department of Energy, grant
DE-FG02-92ER-40704. RE is supported by an NSF Career Award
PHY-0747868.

\section*{References}
\bibliographystyle{h-physrev}
\bibliography{fisher}{}

\appendix

\section{Solutions to the Slow Roll Hierarchy  \label{lowscale}}

\subsubsection*{High-$\epsilon$ 1-Parameter:}
The case where we truncate the flow equations at first order
corresponds to a potential that is quadratic, at least around the
pivot. We can solve the flow equations exactly in this case,
irrespective of the value of $\epsilon$:
\begin{equation}
\epsilon(N) = \frac{\epsilon_0}{2N\epsilon_0+1},
\end{equation}
where $\epsilon_{0}$ is the value of $\epsilon$ at the fiducial
scale (where we have set $N=0$). We find that inflation ends after
\begin{equation}\label{Nendeps}
N_{end} = \frac{1-\epsilon_0}{2\epsilon_0}.
\end{equation}
The value of $\epsilon$ can be tuned to obtain the right amount of
inflation to satisfy the matching condition, equation
(\ref{Nefolds}). However, as we have discussed above, in a one
parameter model, $\epsilon$ is tightly constrained by the tilt,
tensor-scalar ratio and tensor tilt. Any tension between these
quantities will result in the need for non-zero higher order HSR
parameters.

\subsubsection*{Low-$\epsilon$ 1-Parameter:}

See section \ref{loweps} for details of this case.

\subsubsection*{Low-$\epsilon$ 2-Parameter:}

In this situation, the third flow equation is
\begin{equation} \label{flow3}
\frac{d\xi}{dN} = \eta\xi.
\end{equation}
In the limit $\epsilon\ll1$ we can solve equations (\ref{flow2}) and
(\ref{flow3}) to obtain
\begin{eqnarray}\label{etano3}
\eta(N) & = & -\sqrt{2 A_{\xi}}\tan\left[\frac{1}{2}\sqrt{2
A_{\xi}}\left(2 B_{\eta}-N\right)\right], \\
\xi(N) & = & A_{\xi}\tan^{2}\left[\frac{1}{2}\sqrt{2 A_{\xi}}\left(2
B_{\eta}-N\right)\right] + A_{\xi},
\end{eqnarray}
where $A_{\xi} = \xi_{0} - \frac{1}{2}\eta_{0}^{2}$ and $B_{\eta} =
\arctan(-\eta_0/\sqrt{2A_{\xi}})/\sqrt{2A_{\xi}}$. Finally,
\begin{eqnarray} \nonumber
\epsilon(N) & = & C_{\epsilon}\sec\left[\frac{1}{2}\sqrt{2
A_{\xi}}\left(2 B_{\eta}-N\right)\right]^{4},
\end{eqnarray}
where $C_{\epsilon} =
\epsilon_0/\sec^{4}(\sqrt{2A_{\xi}}B_{\eta})$.  We can now solve for where inflation ends (where $\epsilon = 1$) in
this approximation:
\begin{equation}
-N = \frac{2\arccos\left( C_{\epsilon}^{1/4}\right)}{\sqrt{2
A_{\xi}}}-2 B_{\eta}
\end{equation}
\begin{figure}[!t]
\begin{center}
\resizebox{6cm}{!}{\includegraphics{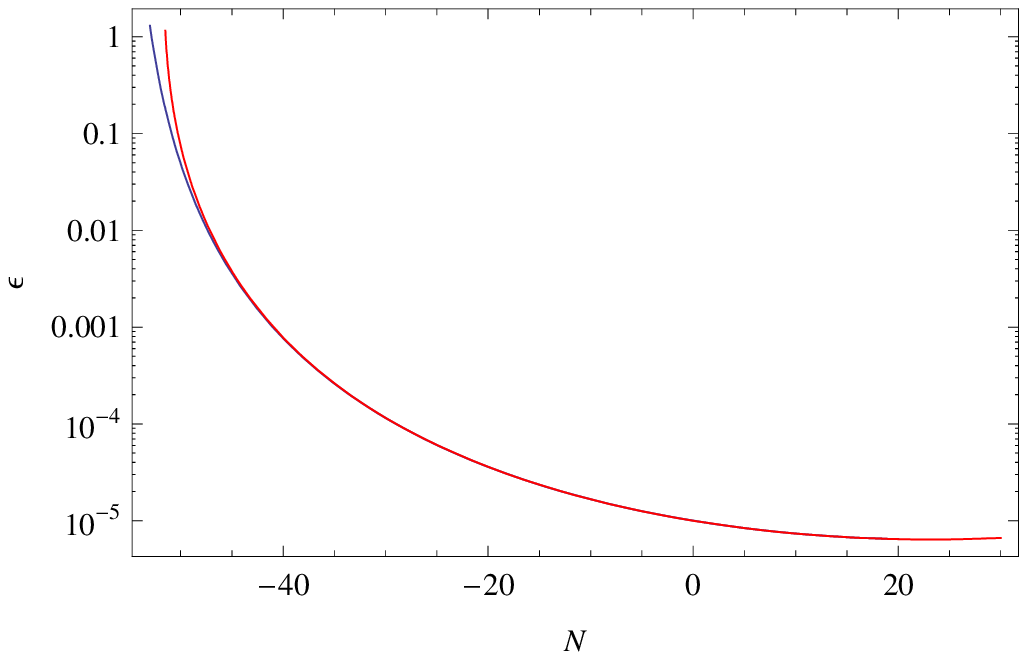}} \hspace{0.13cm}
\resizebox{6cm}{!}{\includegraphics{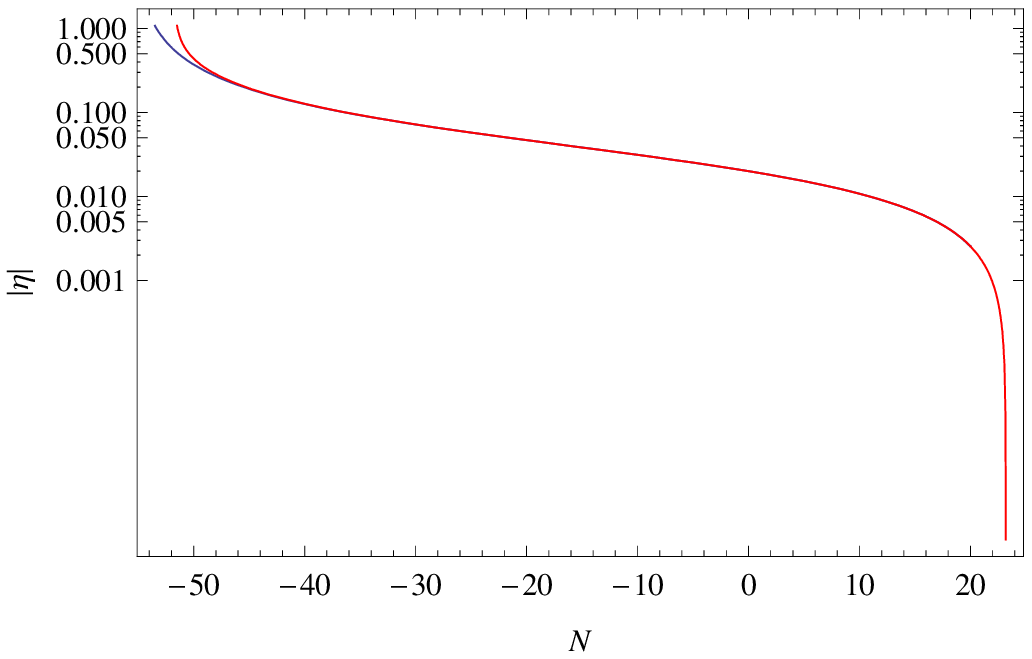}}\\
\resizebox{6cm}{!}{\includegraphics{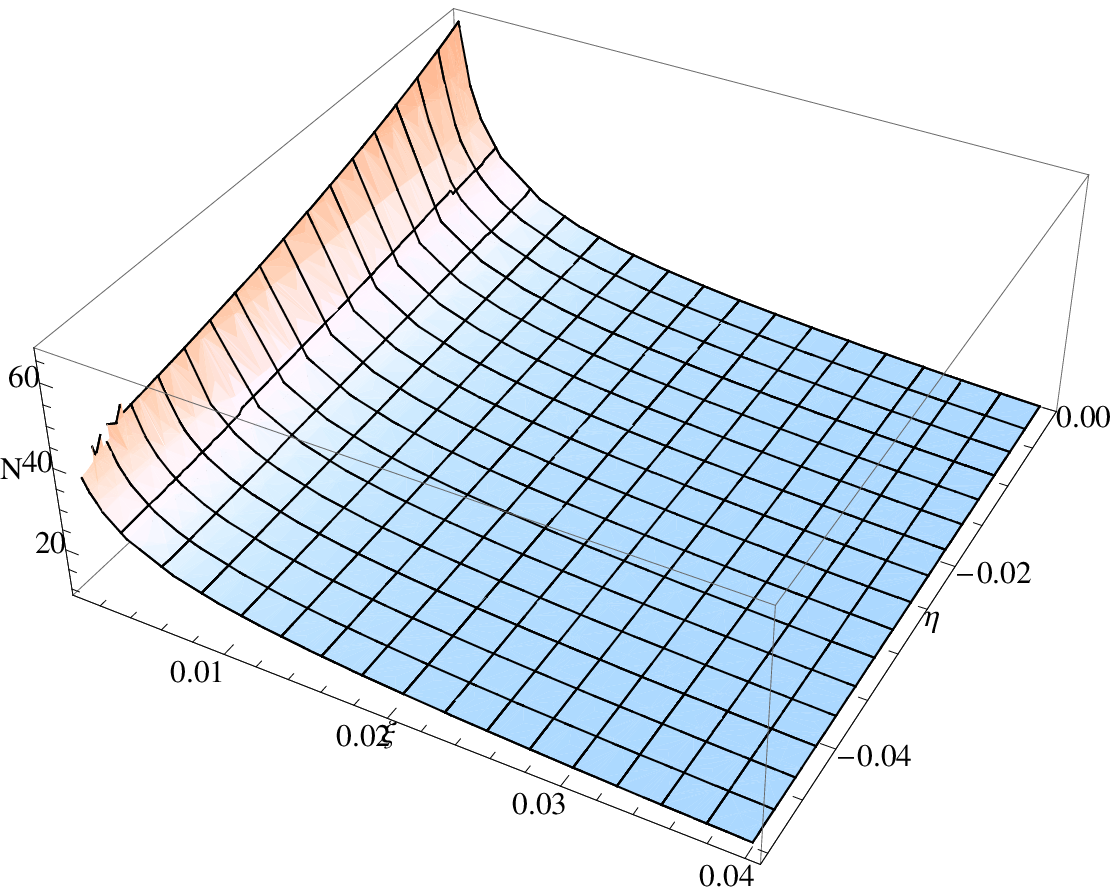}} \hspace{0.13cm}
\resizebox{6cm}{!}{\includegraphics{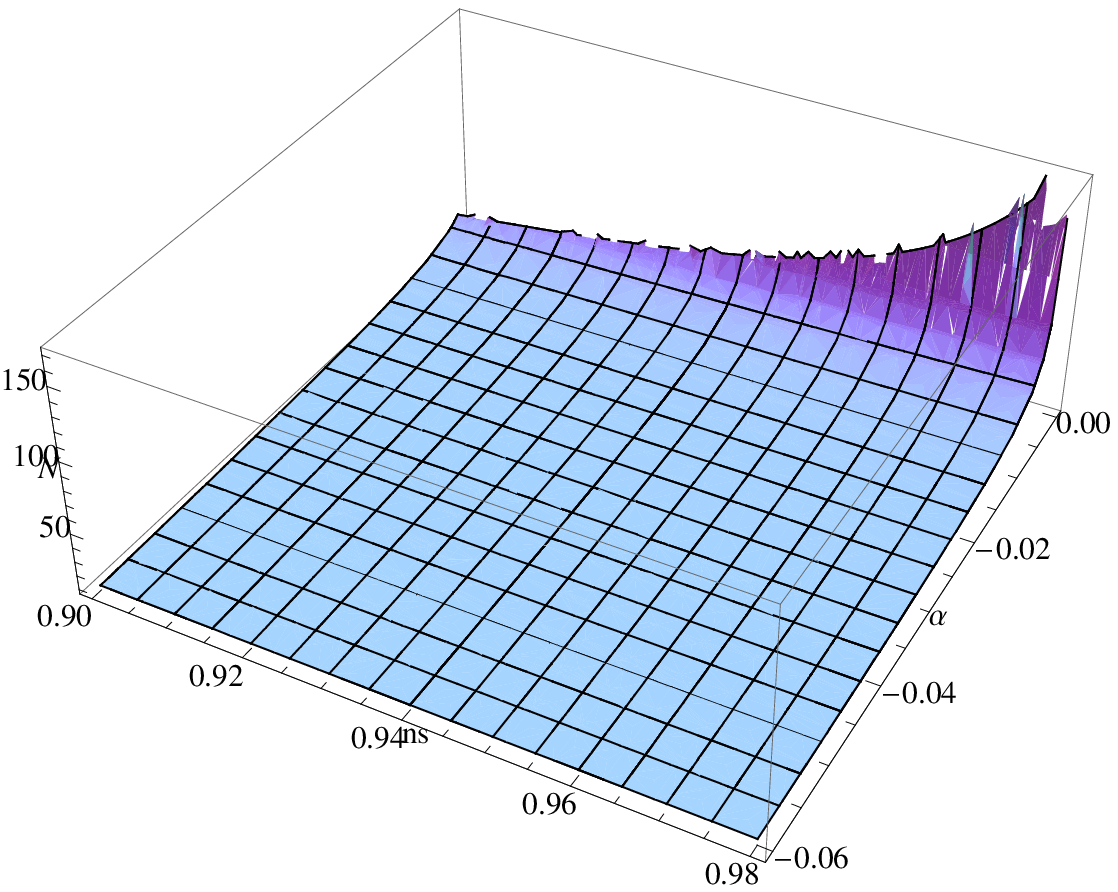}}
\end{center}
\caption{Slow roll hierarchy truncated at third order
(${}^{3}\lambda_{H}=0$) and $\xi = 0.001$. The left panel shows the
evolution of $\epsilon$ while the right panel show $\eta$; the red
curve is the exact result, the blue is the approximation. The lower
left plot shows the amount of inflation as a function of the
parameters $\eta$ and $\xi$, while the lower right plot shows the
same quantity as a function of the spectral indices.}\label{plot2}
\end{figure}
The upper panels of figure \ref{plot2} show the quality of the
approximation, which only gets better as one decreases the value of
$\epsilon$. The lower panels of figure \ref{plot2} shows the number
of e-folds as a function of the slow roll parameters $\eta$ and
$\xi$ and the tilt and running of the scalar spectral index
respectively. The end of inflation in these models is completely
specified by the initial values of the slow roll parameters.

Using  a value of $\xi$ that gives a running anywhere near the centroid of the WMAP data \cite{Spergel:2006hy} leads to a short period of inflation.  If the current
weak evidence for running detected in the WMAP data persists, higher
order slow roll parameters are required to produce enough inflation.
This is consistent with the findings of \cite{Easther:2006tv}.

\subsubsection*{Low-$\epsilon$ 3-Parameter}

\begin{figure*}[!t]
\begin{center}
\resizebox{6cm}{!}{\includegraphics{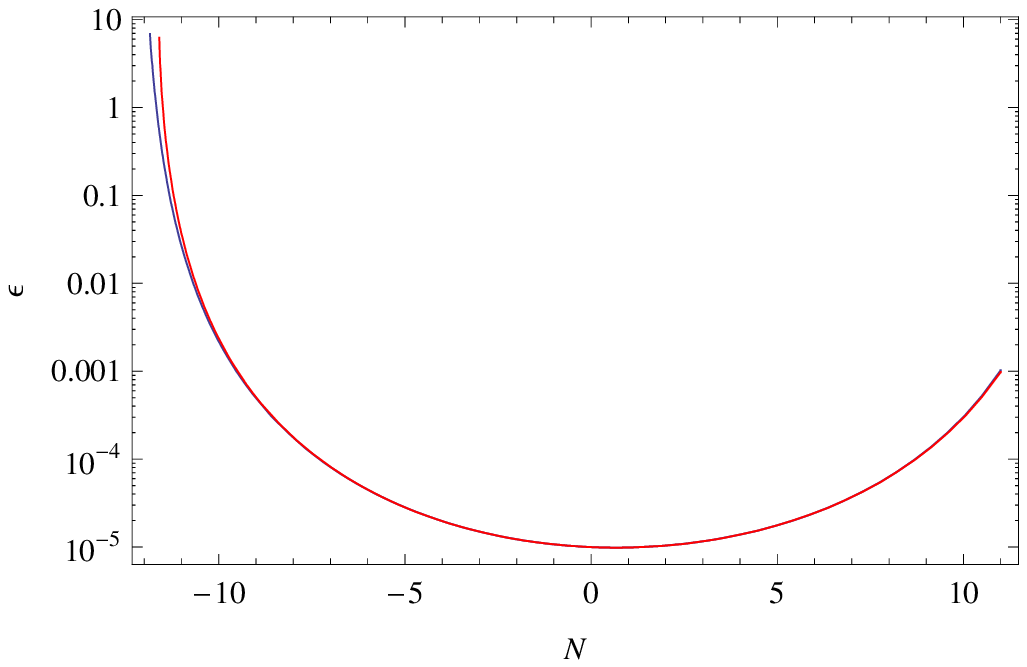}} \hspace{0.13cm}
\resizebox{6cm}{!}{\includegraphics{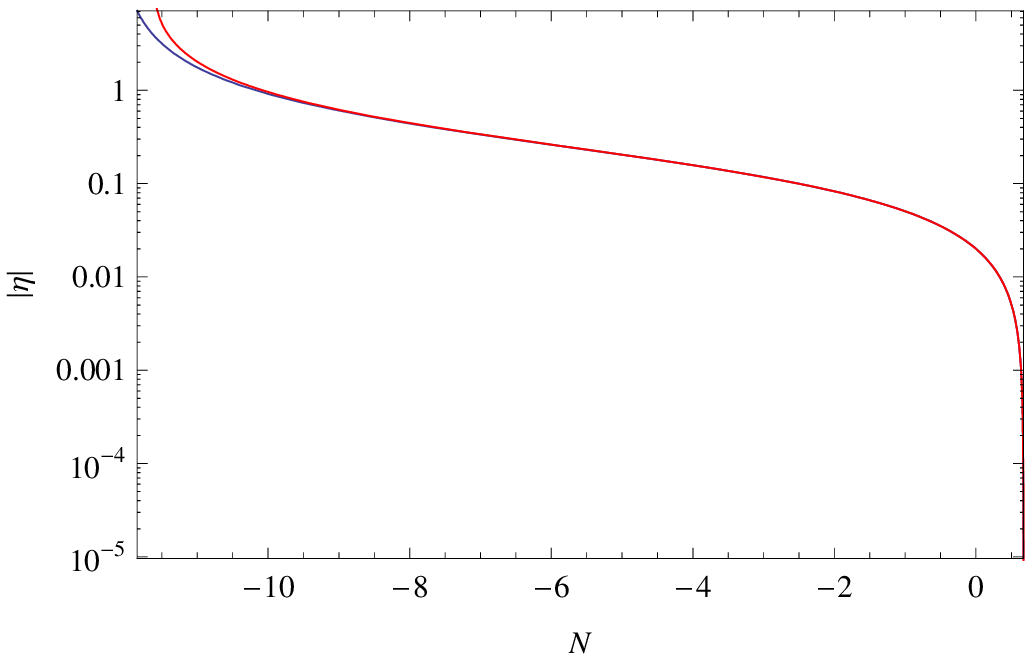}}\\
\end{center}
\caption{Slow roll hierarchy truncated at fourth order
${}^{4}\lambda_{H}=0$. At the pivot, $\eta = -0.02$, $\xi = 0.03$, ${}^{3}\lambda_{H}=-0.0001$ and
$\epsilon=10^{-5}$. Left panel show the evolution of $\epsilon$, the
red curve is the exact result, the blue is the
approximation.}\label{lambnozero}
\end{figure*}

If we include ${}^{3}\lambda_{H}$ there is no obvious analytic solution. However, if
${}^{3}\lambda_{H}$ is small, and noting that $\eta$ is also small and
remains small throughout inflation, we employ a further
approximation:
\begin{equation}\label{3lamHapprox}
\frac{d{}^{3}\lambda_{H}}{dN} \approx 0
\end{equation}
In this limit  $\eta$ obeys Airy's equation, with solution:
\begin{eqnarray}
\eta(N) & =
&-\left(\frac{\pm{}^{3}\lambda_{H}}{2}\right)^{1/3}\left(\frac{
C_{1}\rm{Ai}'\left[x(N)\right] +
\rm{Bi}'\left[x(N)\right]}{C_{1}\rm{Ai}\left[x(N)\right] +
\rm{Bi}\left[x(N)\right]}\right),
\end{eqnarray}
where
\begin{equation}
x(N) =
\mp\left(\frac{\pm{}^{3}\lambda_{H}}{2}\right)^{1/3}\left(N+\frac{A_{\xi}}{{}^{3}\lambda_{H}}\right),
\end{equation}
and $C_{1}$ is a constant given by:
\begin{equation}
C_{1} =-\left( \frac{\rm{Bi}'\left[x(0)\right] +
\left(\frac{\pm{}^{3}\lambda_{H}}{2}\right)^{-1/3}\eta_0
\rm{Bi}\left[x(0)\right]}{\rm{Ai}'\left[x(0)\right]+\left(\frac{\pm{}^{3}\lambda_{H}}{2}\right)
^{-1/3}\eta_0\rm{Ai}\left[x(0) \right]}\right).
\end{equation}
Now we can also solve  for $\epsilon(N)$ to obtain
\begin{equation}
\epsilon(N)= \frac{C_{2}}{\left(C_{1}\rm{Ai}\left[x(N)\right] +
Bi\left[x(N)\right]\right)^{4}}
\end{equation}
where $C_{2}$ is given by:
\begin{equation}
\rm{C_{2}}= \epsilon_0 \left(C_{1}\rm{Ai}\left[x(0)\right]+
Bi\left[x(0)\right]\right)^{4}
\end{equation}
Figure \ref{lambnozero} shows the quality of the approximation, even
for large values of the running, or $\xi$. The value of $\xi$ here
is taken to coincide with a value of $\alpha =-0.06$, i.e. $\xi =
0.03$. Notice how quickly inflation ends; barely 10 e-folds of
inflation are achieved.

These expressions are of limited use in practice. If ${}^{3}\lambda_{H}$ is large enough to prevent $\xi$ from
dominating the dynamics, it must be of the same magnitude as  $\xi$, and equation (\ref{3lamHapprox}) is no longer a good approximation.The effect of including ${}^{3}\lambda_{H}$ depend on its  sign. If ${}^{3}\lambda_{H}<0$, inflation ends more rapidly, for fixed
$\{\epsilon, \eta, \xi\}$. If $0 <{}^{3}\lambda_{H} \ll 1$, we see a slight increase in the number of e-folds. As one increases
${}^{3}\lambda_{H}$, one encounters a critical point is reached where
$\epsilon$ has a maximal value less than unity and then begins to decrease.

\end{document}